\documentclass{elsart} 
\usepackage{graphicx,natbib,amssymb} 
\usepackage{bm}
\journal{Mechanics of Material} 
\begin{document} 
\begin{frontmatter} 
\title{Quasistatic rheology, force transmission and fabric properties of a packing of irregular polyhedral particles} 
\author{E. Az\'ema\corauthref{lmgc_aut} \corauthref{lcpc_auteur}}, 
\ead{emilien.azema@lcpc.fr} 
\author{F. Radjai\corauthref{lmgc_aut}},
\ead{radjai@lmgc.univ-montp2.fr} 
\author{G. Saussine\corauthref{sncf_aut}}
\ead{gilles.saussine@sncf.fr}
\address[lmgc_aut]{Laboratoire de M\'ecanique et G\'enie Civil, Universit\'e Montpellier 2, Place Eug\`ene Bataillon, 34095 Montpellier cedex 05} 
\address[lcpc_auteur]{Present address : Laboratoire Central des Ponts et Chauss\'ees, D\'emarches Durables en G\'enie Civil, 44341 Bouguenais cedex
(Site de Nantes)} 
\address[sncf_aut]{Innovation and Research Departement of SNCF, 45 rue de Londres, 75379 Paris Cedex 08} 
\begin{abstract} 
By means of contact dynamics simulations, we investigate  
a dense packing composed of polyhedral particles under quasistatic shearing.   
The effect of particle shape is analyzed by comparing the 
polyhedra packing with a packing of similar characteristics except for 
the spherical shape of the particles. The polyhedra packing shows higher shear stress 
and dilatancy but similar stress-dilatancy relation compared to the sphere packing. 
A harmonic approximation of granular fabric is presented in terms of branch vectors 
(connecting particle centers) and contact force components along and perpendicular 
to the branch vectors. It is found that  the 
origin of enhanced shear strength of the polyhedra packing lies in its  
higher force anisotropy with respect to the sphere packing which  
has a higher fabric anisotropy. Various contact types 
(face-vertex, face-face, etc) contribute differently to force transmission 
and fabric anisotropy. In particular, most face-face contacts  belong 
to strong force chains along the major principal stress direction 
whereas vertex-face contacts are correlated with weak forces 
and oriented on average along the minor principal stress direction 
in steady shearing.      
\end{abstract} 
\end{frontmatter} 

%-----------------------------------------------------------------------------------------------------

\section{Introduction}
\label{sec:int}
Many recent numerical studies of granular media 
deal with model systems composed of spherical particles. 
The use of simplified particle shapes and contact interactions is 
needed in order to focus on the collective behavior of particles 
which is at the origin of many specific properties of granular 
materials. On the other hand, the numerical treatment of complex particle 
shapes by discrete element methods was until very recently  out of reach due to  
demanding computational resources.  There 
is presently, however, considerable scope for the numerical investigation of 
complex granular packings. This is not only due to 
available computer power and memory but also because during 
more than two decades of intense research in this field, many fundamental aspects of  
granular media 
have already been established for simplified particle shapes. 
In particular, various microscopic features such as 
fabric anisotropy (\cite{Kruyt1996,Bathurst1988,Rothenburg1989, Radjai1998, Kruyt2004}), 
force transmission (\cite{Liu1995a,Radjai1996,Coppersmith1996,Mueth1998a,Lovol1999, Bardenhagen2000, Antony2001,Silbert2002,Metzger2004a,Majmudar2005}) 
and friction mobilization (\cite{Radjai1998,Staron2005})
have been analyzed for circular particles (in 2D) and  
spheres (in 3D). Hence, a recurrent issue today is how 
robust these findings are with respect to particle shape 
(\cite{Ouadfel2001,Antony2004,Cambou2004,Nouguier-Lehon2003,Alonso-Marroquin2002,Pena2005,Pena2006,Pena2006a,Azema2007}).  

The issue of shape effect opens actually the door 
to a vast and substantial scientific domain given a 
multitude of potential particle morphologies.  Several 
well-known examples are elongated and platy shapes (occurring  
in biomaterials and pharmaceutical applications), angular and facetted 
shapes (occurring in geomaterials) and nonconvex 
shapes (occurring in sintered powders). 
The macroscopic shear behavior is considerably influenced by particle 
shape. Rounded particles enhance flowability whereas   
angular shape is susceptible to improve shear strength, a factor of vital importance to  
civil-engineering applications (\cite{Nouguier-Lehon2003}). 
In many engineering applications  the particle shapes need to be optimized in order 
to increase performance (\cite{Markland1981,Wu2000,Lim2005,Saussine2006,Lobo-Guerrero2006,Lu2007}). 

In this paper, we employ the contact dynamics method to investigate 
the slow shear behavior of granular media composed of polyhedral particles. 
The facetted shapes give rise to a rich microstructure where the particles 
touch at their faces, edges and vertices. Face-face contacts are expected to 
play a major role in force transmission and statics of polyhedra by accommodating  
long force chains that are basically unstable in a packing composed 
of spheres. In order to isolate the effects arising from particle shape,  
the data from the polyhedra packing will be compared with a packing 
of spherical particles that, apart from the particle shape, is identical 
in all respects (preparation, friction coefficients, particle size distribution) 
to the polyhedra packing. Both packings are subjected to monotonous triaxial 
compression.  

The numerical procedures will be presented with a brief technical introduction 
to the detection and treatment of contacts between polyhedra in the 
framework of the contact dynamics method.   
We will consider the stress-strain and volume-change behavior. 
The harmonic approximation of the fabric and additive decomposition of 
the stress tensor into fabric and force anisotropies will be presented 
in detail. This will allow us to assess in clear terms the origins of shear strength in 
the polyhedra packing from 
fabric and force anisotropies in comparison to the sphere packing. 
The probability density functions of normal forces will be studied and 
compared between the two assemblies.   
Finally, we will focus on the contact networks of polyhedral particles and 
the role played by different contact categories with respect to 
force transmission.

%-----------------------------------------------------------------------------------------------------
\section{Numerical method}
\label{sec:num}
In this section we briefly introduce the contact dynamics (CD) method
with polyhedral particles and the numerical procedures used 
for sample preparation.   

\subsection{Contact dynamics method with polyhedra} 

The CD method is based on implicit time integration 
and nonsmooth formulation of
mutual exclusion and dry friction between particles  (\cite{jean1992,Moreau1994,Radjai1999,DUBOIS2003}).
The equations of motion are formulated 
as differential inclusions in which
velocity jumps replace accelerations (\cite{Moreau1994}). 
The unilateral contact interactions and Coulomb friction law are represented as set-valued
force laws. The implementation of the time-stepping scheme requires the 
geometrical description of each potential 
contact in terms of contact position and its normal unit vector.  

At each time step, all kinematic constraints implied by enduring  contacts  
are {\it   simultaneously} taken into account together with  the equations 
of motion in order to determine all velocities and contact forces in the system. 
This problem is solved by an iterative process  pertaining to the 
non-linear Gauss-Seidel method that consists
of solving a single contact problem, with other contact forces being treated 
as known, and iteratively 
updating the forces until a given convergence criterion is achieved. 
The method is thus able to deal properly with the {\it nonlocal} character 
of the momentum transfers resulting from the impenetrability  of the  
particles and friction law. 

The CD method is unconditionally stable due to its inherent 
implicit time integration method. 
The uniqueness of the solution at each time step is not guaranteed 
for perfectly rigid particles. However, by initializing each step 
with the forces calculated  in the preceding step, the variability of admissible 
solutions shrinks to the numerical resolution. In the discrete element 
methods based on molecular dynamics, this ``force history" is,  by construction, included  
in the particle positions.

The treatment of a contact interaction between two particles requires 
the identification of the contact zone and a  ``common plane". 
For rigid particles it is possible to define this contact zone by a finite set of points.
Before applying the contact detection algorithm between a pair of particles of  
irregular shapes, a ``bounding box" method is used to compute a list 
of particle pairs potentially in contact.
Then, for each pair, the first step is to determine if an overlap exists through a 3D 
extension of the ``shadow overlap method" (\cite{Saussineoctober2004,DUBOIS2003}).
 Several algorithms exist for overlap determination between convex polyhedra 
 (\cite{Cundall1979,Cundall1988,nezami2004,Nezami2006,DUBOIS2003,Saussineoctober2004,Saussine2006,Perales2007}). 
 When an overlap occurs, 
 the contact plane is determined by 
computing the intersection between the two particles.

\begin{figure}
\centering
\includegraphics[width=6cm]{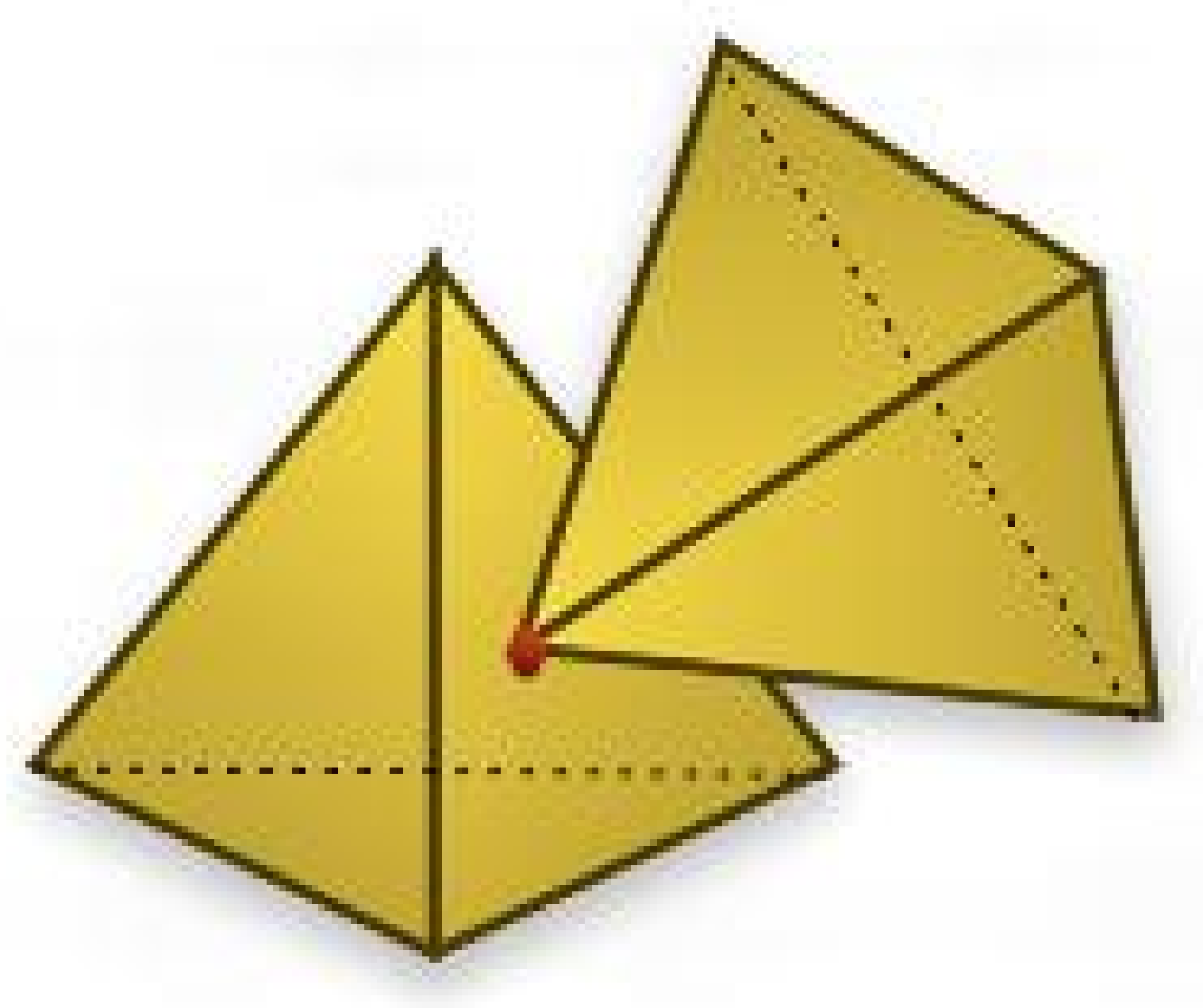}
\includegraphics[width=6cm]{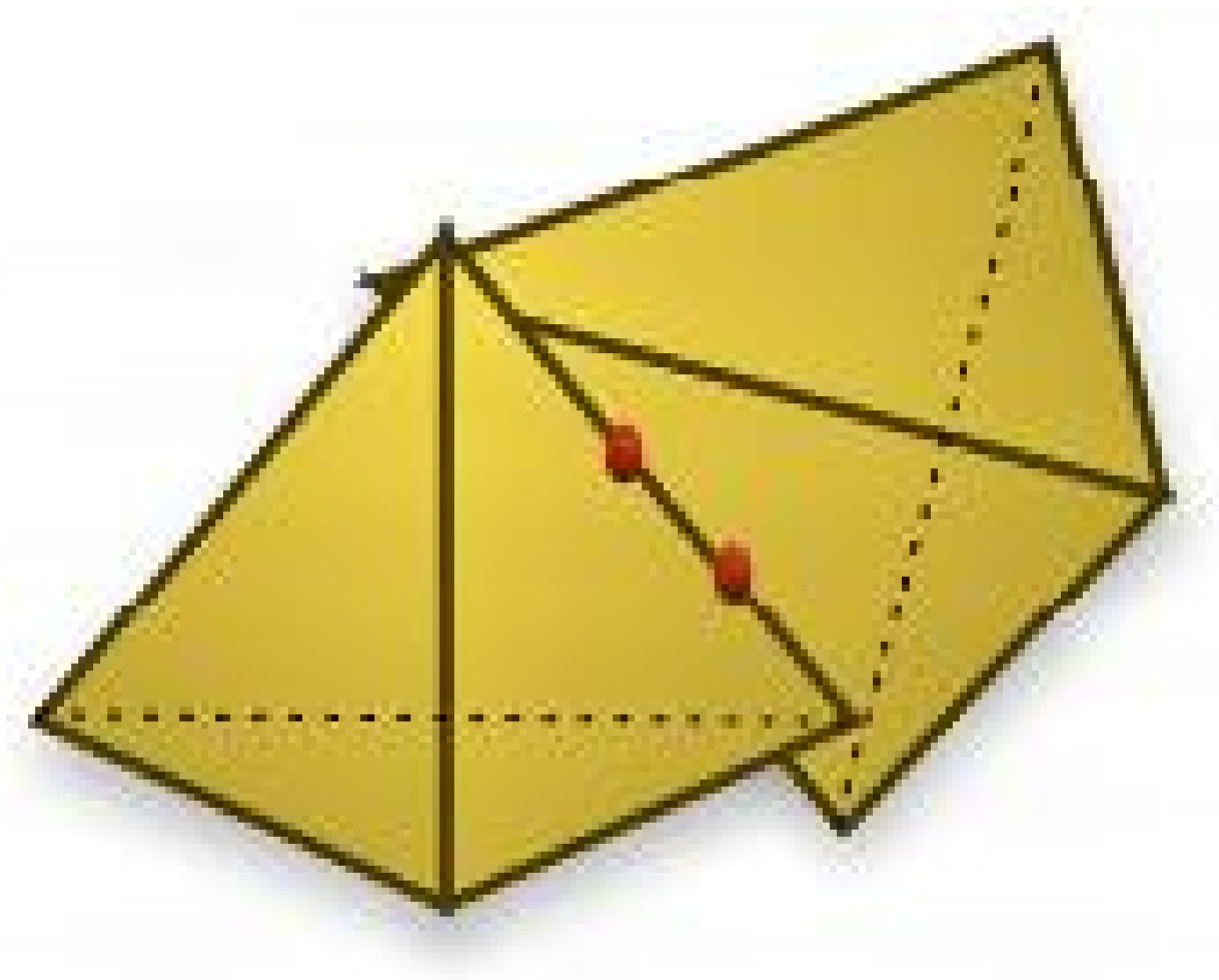}
\includegraphics[width=6cm]{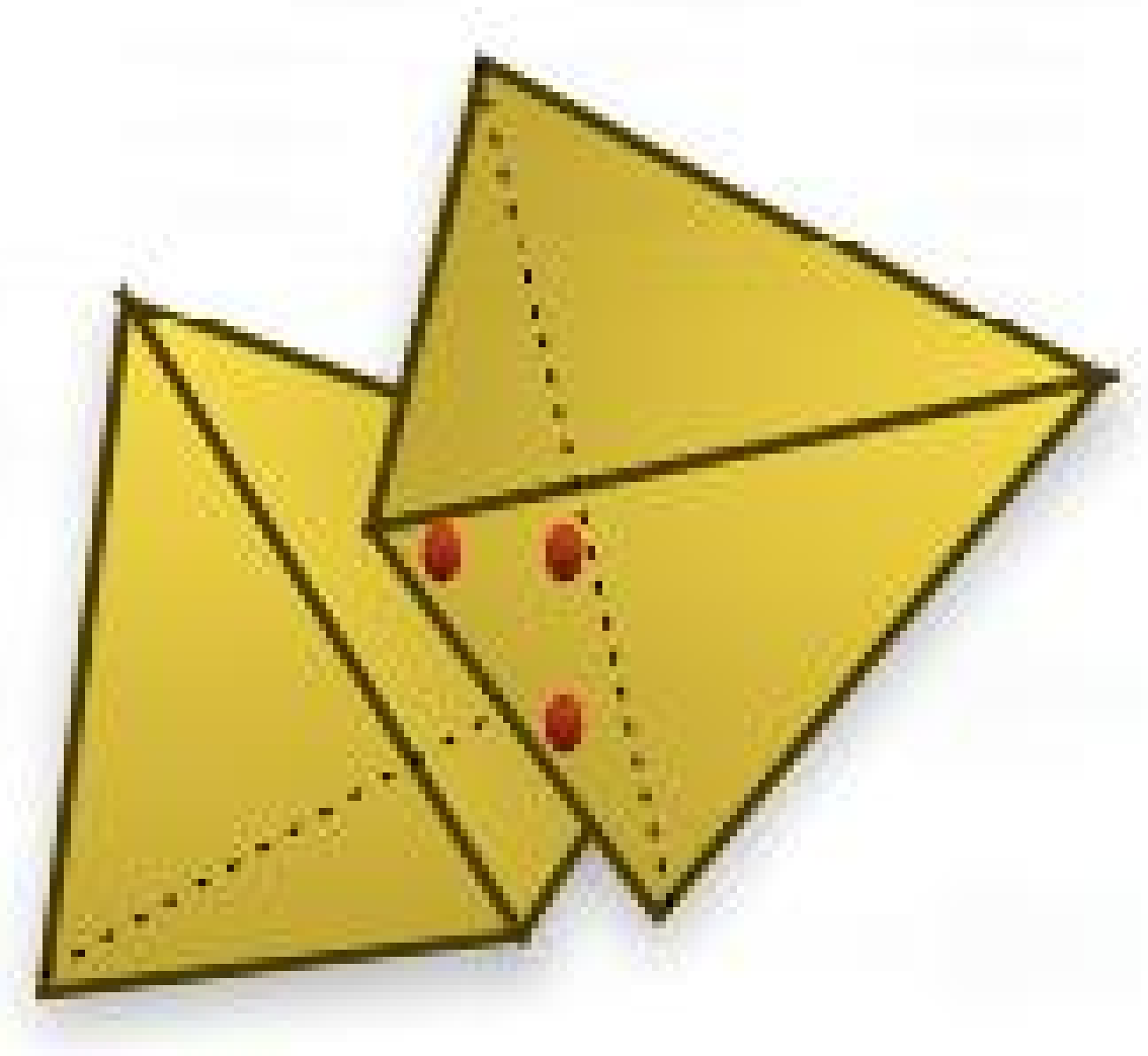}
\caption{Different types of contacts between two polyhedra.}
\label{fig:01}    
\end{figure}

The contacts between polyhedral particles belong to different 
categories, namely face-face, edge-face, vertex-face, edge-edge, vertex-vertex, vertex-edge; see Fig. \ref{fig:01}. The vertex-vertex and vertex-edge 
contacts are practically absent. In all cases, we determine 
one, two or three contact points which provide a good description of 
the contact zone.
In this paper, the vertex-edge and edge-edge contacts are 
referred to  as ``simple" contacts whereas the edge-face and 
face-face contacts are treated as 
``double" and ``triple" contacts since their representation 
involves 2 and 3 distinct points on the common plane, respectively. 

For our simulations, we used the LMGC90 which is a multipurpose software developed in Montpellier, capable of modeling a collection of deformable 
or undeformable particles of various shapes 
by different resolution algorithms (\cite{DUBOIS2003}).

\subsection{Sample preparation}

\begin{figure}
\centering
\includegraphics[width=8cm]{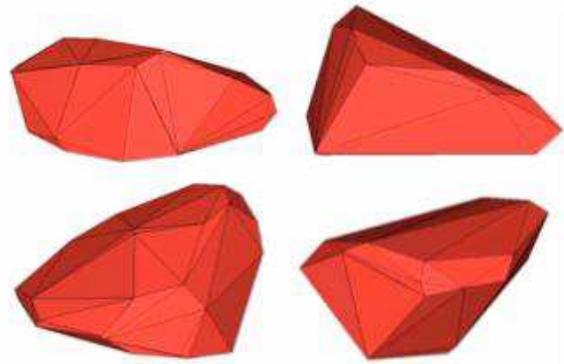}
\caption{Examples of polyhedra used in the simulations.}
\label{grains_ballast}    
\end{figure}

\begin{figure}
\includegraphics[width=8cm]{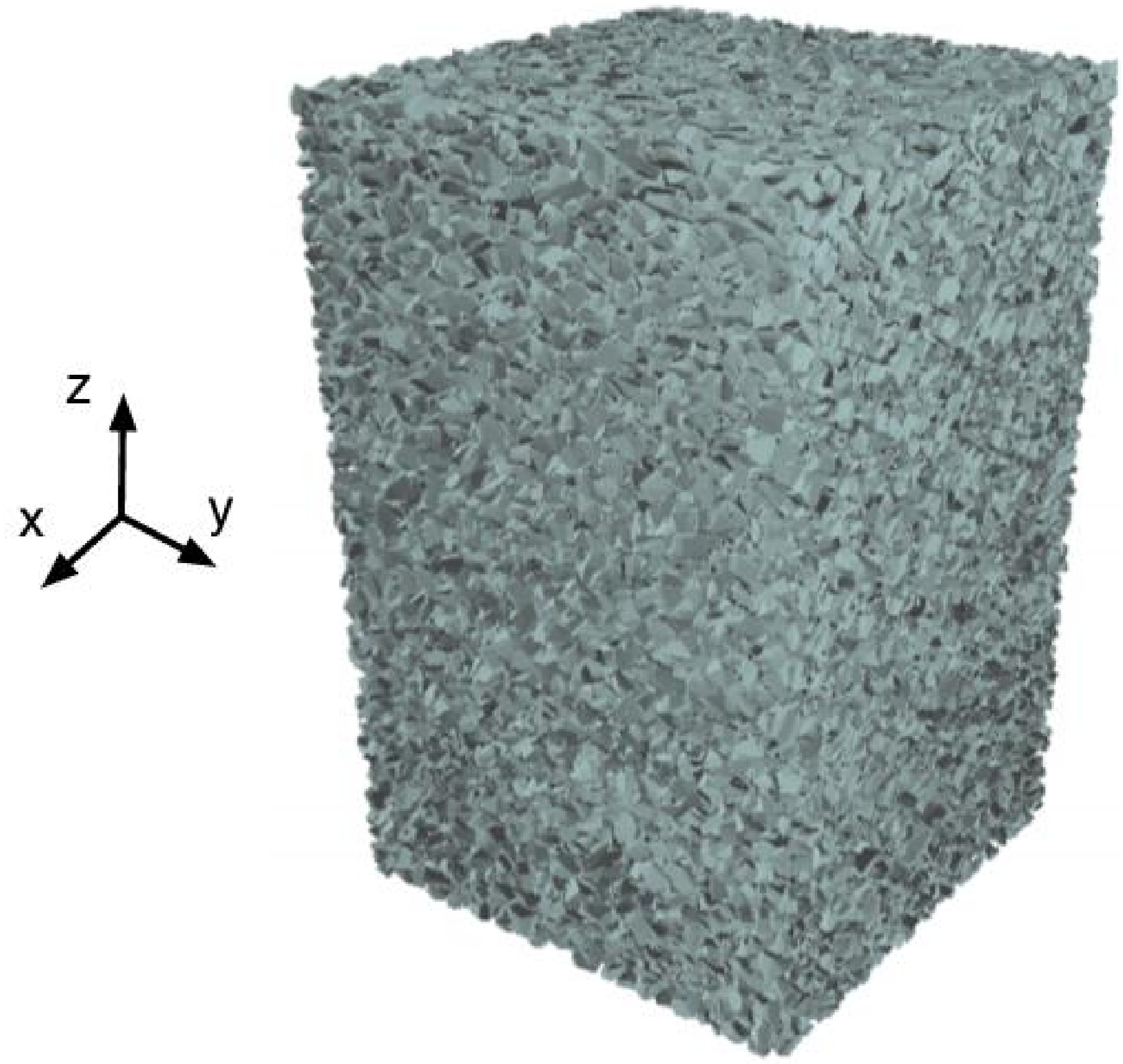}
\includegraphics[width=7cm]{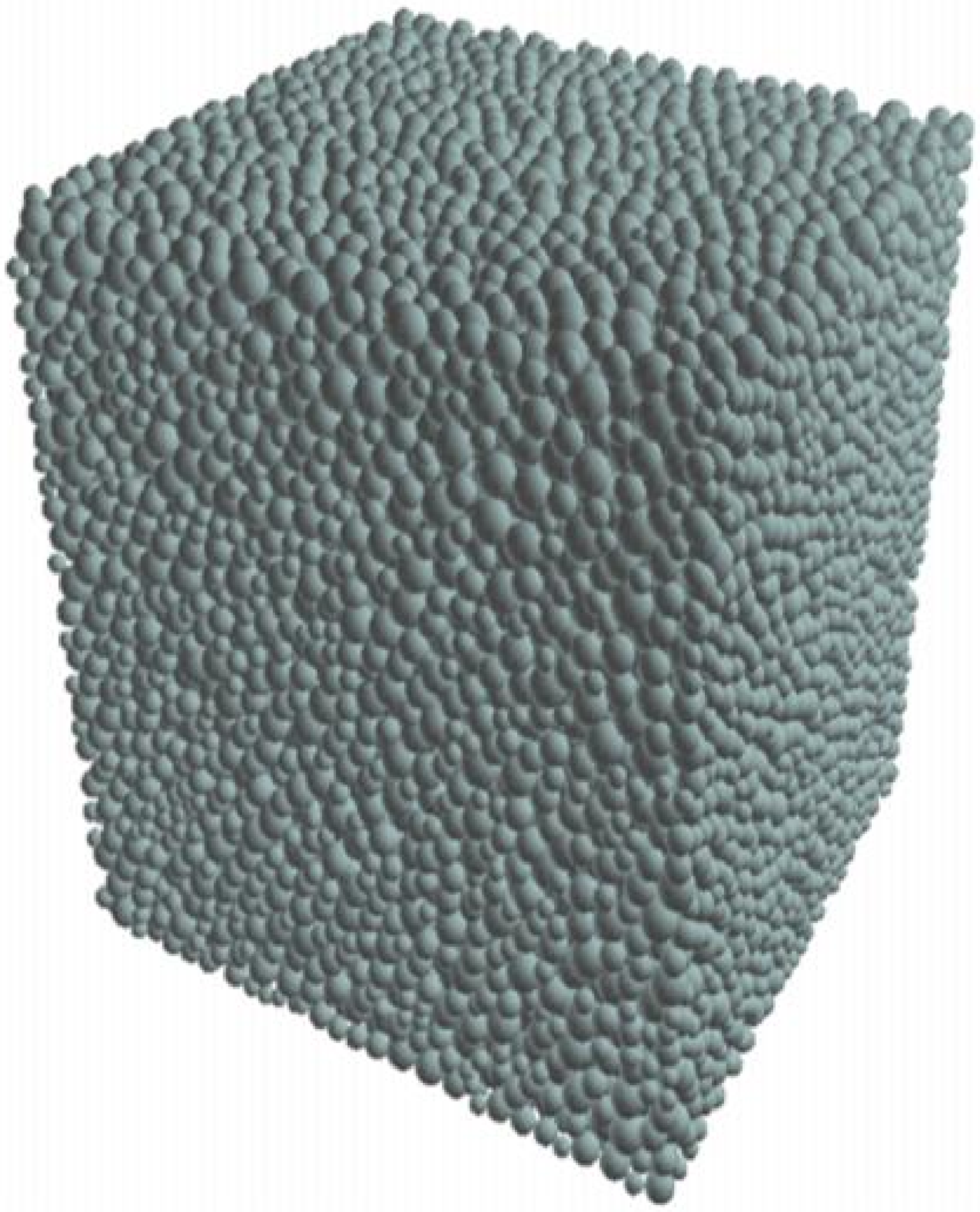}
\caption{Snapshots of the two packings S1 (polyhedra) and S2 (spheres). The 
walls are not shown}
\label{SamplesPOLYH_SPHER}    
\end{figure}

We generate two numerical samples. The first sample (S1) is composed of 
36933 polyhedra. The particle shape are taken from a library of 1000 digitalised ballast grains provided by the French Railway
Company SNCF.
Each particle has at most 
70 faces and 37 vertices and at least 12 faces and 8 vertices. 
Fig. \ref{grains_ballast} shows several examples of the polyhedral 
particles used in the simulations.   
The size of a particle  is defined as two times the largest distance between the 
barycenter and the vertices of the particle, to which we will refer as ``diameter" below. 
We used the following size distribution:  
50\% of diameter $d_{min}=2.5$ cm, 34\% of diameter $3.75$ cm, 
16\% of diameter $d_{max}=5$ cm. This distribution represents  an approximation of 
that of railway ballast grains. The sample contains $7.1 \ 10^5$ vertices and more than 
$10^6$ faces, the average numbers being 20 and 35, respectively.  
The second sample (S2) is composed of 19998 spheres with exactly the same 
size distribution as in S1. Fig. \ref{SamplesPOLYH_SPHER} shows a snapshot of the two samples in equilibrium state after deposition and isotropic compression 
under a constant stress of $\sigma_0=10^4$ Pa  in a rectangular box at zero gravity. 

The coefficient of friction is 0.5 between the particles in both samples and 0 
with the walls. The normal and tangential coefficients of restitution are 0.   
The zero restitution simplifies the deposition and compaction process by enhancing 
dissipation during dynamics rearrangements.  
The initial value of the solid fraction is $\rho \simeq 0.6$ in both samples. 
Both samples have a nearly square bottom of side such that $L\approx l$ 
and an aspect ratio $H/L \simeq 2$, where $H$ is the height. The initial
configuration is defined by $H_0\simeq 30 D_{M}$ for S1 and  for S2 with $D_{M}$ the mean diameter.

The isotropic samples are subjected to vertical compression by imposing a 
constant downward velocity of $10$ cm/s on the upper wall and a constant 
confining stress  $\sigma_2=\sigma_3=\sigma_0$ on the lateral walls. 
Each simulation is stopped for a vertical deformation of $30\%$. 
The time step was  $2.10^{- 4}$ s. 
The CPU time was  $2 \ 10^{- 3}$ s for S1 and $1 \ 10^{- 3}$ s for S2, 
per particle and per time step on an Apple G5 computer. 
The deformation process can be considered to be quasistatic in view of the 
weak kinetic energy injected into the samples compared to the static pressure. 
This can be expressed more generally through the inertial number defined as (\cite{GDR-MiDi2004}): 
\begin{equation}
I=\dot \varepsilon \sqrt{\frac{m}{dp}},
\label{eq3}
\end{equation}
where $\dot \varepsilon=\dot H /H$ is the vertical strain rate, $m$ is the total 
mass, $p$ is the mean pressure and $d$ is the mean particle diameter.  
In our simulations, we have $I \simeq 10^{-3}$, corresponding 
to the quasistatic limit. 

%-----------------------------------------------------------------------------------------------------

\section{Stress-strain behavior}
\label{sec:str}
In this section, we compare the stress-strain and volume-change behavior 
between the packings of polyhedra (packing S1) and spheres (packing S2). 
The stress and strain variables are defined from numerical data.     
For the estimation of the stress tensor, we use the "tensorial 
moment" ${\bm M}^i$ of each particle i defined by (\cite{Moreau1997,Staron2005}):
\begin{equation}
M^i_{\alpha \beta} = \sum_{c \in i} f_{\alpha}^c r_{\beta}^c,
\label{eq:M}
\end{equation}
where  $f_{\alpha}^c$ is the $\alpha$ component of the force exerted on 
particle i at the contact c, $r_{\beta}^c$ is the $\beta$ component 
of the position vector of the same contact c, and the summation 
runs over all contact neighbors of particle i 
(noted briefly by $c \in i$).

It can be shown that the tensorial moment of a collection of 
rigid particles is the sum of the tensorial moments of 
individual particles (\cite{Moreau1997}). 
The stress tensor ${\bm \sigma}$ for a packing of volume $V$  
is simply given by (\cite{Moreau1997,Staron2005}): 
\begin{equation}
{\bm \sigma } = \frac{1}{V} \sum_{i \in V} {\bm M}^i =   \frac{1}{V}  \sum_{c \in V} f_{\alpha}^c \ell_{\beta}^c, 
\label{eq:M}
\end{equation}
where ${\bm \ell}^c$ is the branch vector joining the centers of 
the two touching particles at the 
contact $c$. Remark that the first summation runs over all 
particles whereas the second summation 
involves the contacts, each contact appearing only once.

Under triaxial conditions with vertical compression, we have 
$\sigma_1 \geq \sigma_2=\sigma_3$, where the $\sigma_\alpha$ are 
the stress principal values.  
Using the Cambridge representation, we define the mean stress $p$ and
stress deviator $q$ by (\cite{Airey1988}) : 
\begin{eqnarray}
p &=&  \frac{1}{3}(\sigma_1 + \sigma_2 + \sigma_3), \\
q &=& \frac{1}{3}(\sigma_1 - \sigma_3).
\label{eq:q}
\end{eqnarray}
For our system of perfectly rigid particles,  the stress state is characterized by 
the mean stress $p$ and the normalized shear stress $q/p$.    

The cumulative strain components $\varepsilon_\alpha$ are defined by 
\begin{eqnarray}
\centering
\varepsilon_1&=&\int_{H_0}^H \frac{dH'}{H'} = \ln \left( 1+ \frac{\Delta H}{H_0} \right),\\
\varepsilon_2&=&\int_{L_0}^L \frac{dL'}{L'} = \ln \left( 1+ \frac{\Delta L}{L_0} \right),\\
\varepsilon_3&=&\int_{l_0}^l \frac{dl'}{l'} = \ln \left( 1+ \frac{\Delta l}{l_0} \right),
\label{eq:vara}
\end{eqnarray}
where $H_0$, $l_0$ and $L_0$ are the initial height, width and length of the 
simulation box, respectively and   $\Delta H = H_0 - H$,  $\Delta l = l_0 - l$ and  $\Delta L = L_0 - L$ are the corresponding cumulative displacements. The 
volumetric strain is given by 
\begin{equation}
\varepsilon_p = \int_{V_0}^V \frac{dV'}{V'} = \ln \left( 1+ \frac{\Delta V}{V_0} \right),
\label{eq:varV}
\end{equation}   
where $V_0$ is the initial volume  and   
$\Delta V = V - V_0$ is the total volume change. 
The cumulative shear strain is defined by 
\begin{equation}
\varepsilon_q \equiv \varepsilon_1-\varepsilon_2.
\label{eq:qV}
\end{equation}

\begin{figure}
\centering
\includegraphics[width=8cm]{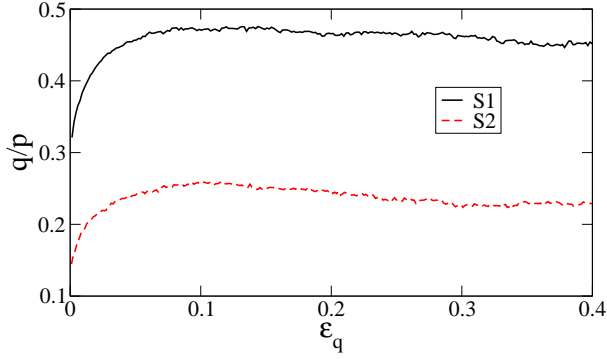}
\caption{The strength parameter $q/p$ as a function of shear strain $\varepsilon_q$
for the polyhedra packing S1 and sphere packing S2.}
\label{qp_all_3D}    
\end{figure}

Figure \ref{qp_all_3D} displays the evolution of $q/p$ for the packings S1 and S2 
as a function of $\varepsilon_q$. 
For both packings, we observe a classical behavior characterized by a 
hardening behavior followed by (slight) softening and a stress 
plateau corresponding to the 
critical state of soil mechanics (\cite{Mitchell2005}). 
The critical-state strength in the case of polyhedra ($\simeq 0.46$) 
is twice as high as that of spheres ($ \simeq 0.23$). 
This implies that the polyhedra packing has a higher angle of 
internal friction $\varphi$ defined by 
\begin{equation}
\centering
\sin \varphi = \frac{3q}{2p+q}.
\end{equation}
At the critical state, we  have $\varphi = \varphi_0=34^\circ$ for S1 and 
$\varphi_0=18^\circ$ for S2. 

\begin{figure}
\centering
\includegraphics[width=8cm]{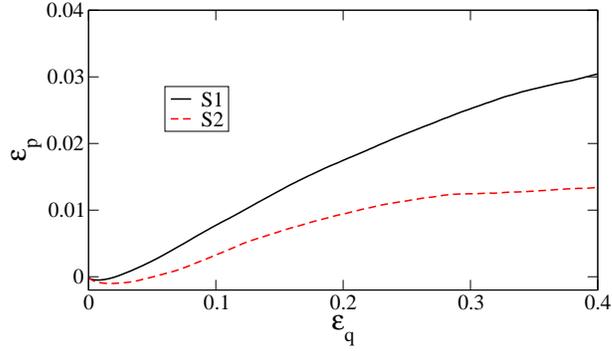}
\caption{The volume change  $\varepsilon_p$ as a function of 
shear strain  $\varepsilon_q$ for the packings S1 et S2.}
\label{dilatance_all_3D}    
\end{figure}

Figure  \ref{dilatance_all_3D} shows the volumetric strain   
$\varepsilon_p$ as a function of shear strain  $\varepsilon_q$ in S1 and S2. 
In both packings, we observe an early compaction slightly larger in S2 than 
in S1.  The subsequent dilation is lower in S2 and the critical state with isochoric 
deformation is reached at   $\varepsilon_q = 0.3$. Dilation in S1 continues 
with a decreasing rate of volume change but  the isochoric plateau is not fully reached. 
The dilatancy can be expressed in terms of the dilation angle $\psi$ defined by     
\begin{equation}
\centering
\sin \psi =  \frac{\varepsilon_p}{\varepsilon_q}.
\end{equation}
We have $\psi \simeq 5^\circ$ for S1 and $\psi \simeq 2.5^\circ$ for S2 
at the stress peak state. 
 
\begin{figure}
\centering
\includegraphics[width=8cm]{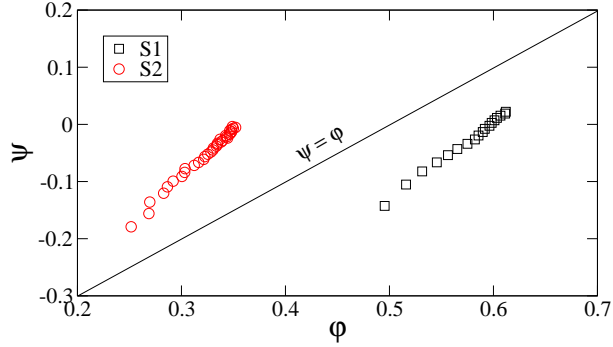}
\caption{The stress-dilatancy diagram representing the relation between the internal angle of friction and the dilation angle  for polyhedra and spheres.}
\label{phi_psi_3D}    
\end{figure}

The variation of $\psi$ versus  $\varphi$, a sort of 
stress-dilatancy diagram (\cite{Wood1990}), is displayed in Fig. \ref{phi_psi_3D} for 
polyhedra and spheres. For both packings, we have  
\begin{equation}
\centering
\varphi \simeq k \psi + \varphi_0, 
\label{eq:sd}
\end{equation}
where $k$ is a constant slightly smaller than 1 in both packings. This correlation between 
dilatancy and shear stress during stress-strain transients is a consequence 
of energy balance. The mechanical work performed on  the system is partially 
dissipated in contact interactions and partially used in volume change (\cite{Radjai2004}).   
Several stress-dilatancy relations have been proposed 
as flow rules for plastic deformations of granular media (\cite{Wood1990}). The 
relation (\ref{eq:sd}) associates the peak state to the largest positive value of dilatancy and 
the critical state to zero dilatancy. It shows the ``non associated"  character of the flow rule 
in granular media (an associated flow rule implying $\varphi = \psi$).   

%-----------------------------------------------------------------------------------------------------

\section{Harmonic representation of the fabric}
\label{sec:har}

The expression of stress tensor in Eq. (\ref{eq:M}) 
is an arithmetic mean involving  
the branch vectors and contact forces. Hence,  
in order to analyze the shear strength properties of 
the polyhedra packing compared to the sphere packing,  
we need a statistical description of the granular 
microstructure (texture or fabric) and 
force transmission. 

In the presence of steric exclusions, the 
granular microstructure is highly disordered at the particle scale 
(\cite{Troadec2002a,Troadec2002}). Since mechanical interactions are governed by 
contact and friction, the relevant descriptors of the microstructure 
are related to the contact network. At the lowest order, the contact network is 
characterized by the coordination number $z$ which describes the compactness of 
a packing. This is a crude scalar information in view of the complex  arrangement of 
the particles, but it is well-known that the compactness, generally expressed in terms 
of the solid fraction, controls the stress-strain behavior under monotonous shearing. 
Let us remark here that double and triple contact types (see section \ref{sec:num}) 
are counted as {\em single} contacts for the coordination number although they 
are represented by two and three contact points, respectively, in the numerical 
treatment of interactions between polyhedra.          

\begin{figure}
\centering
\includegraphics[width=8cm]{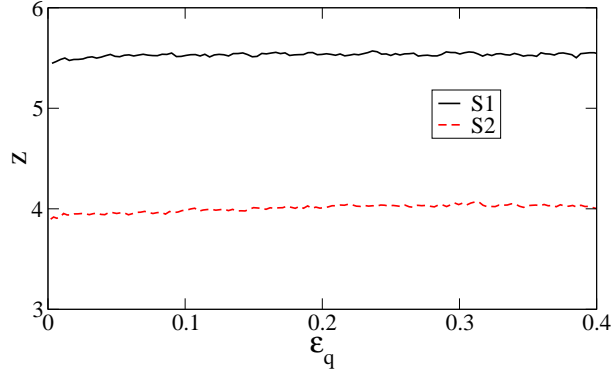}
\caption{Evolution of the coordination number $z$ as a function 
of the cumulative shear strain 
$\varepsilon_q$ for polyhedra (S1) and spheres (S2)}
\label{z_3D}    
\end{figure}

\begin{figure}
\centering
\includegraphics[width=8cm]{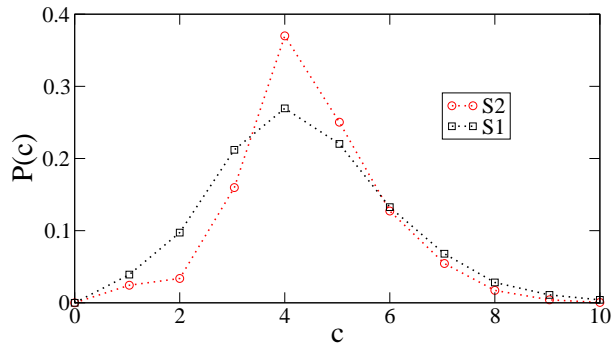}
\caption{The connectivity $P(c)$ of the contact network for the packings S1 and S2.}
\label{pc_3D}    
\end{figure}

The evolution of $z$ for polyhedra and spheres is shown in Fig. \ref{z_3D}  
as a function of $\varepsilon_q$. It is remarkable that   
$z$ is nearly constant in spite of the overall dilation in both packings. 
We have  $ z\simeq 5.5$ for polyhedra and  $z \simeq 4$ for spheres. 
The connectivity of the contact network can
be characterized in more detail by the 
fraction $P(c)$ of particles with exactly $c$ contact neighbors. 
The coordination number is the mean value of $c$ : $z= \sum_c c P(c)$. 
Fig.  \ref{pc_3D} shows $P(c)$ for S1 and S2 in the critical state. 
 The distribution is broader in S1 than in S2. This shows the wider range 
 of potential equilibrium states in the polyhedra packing.    
 For both packings, we observe a peak centered on $c=4$ with a 
 higher probability for S2.

Since the shear stress corresponds to
the deviation of stress components from the mean stress $p$  
along different space directions, the coordination number $z$ as a scalar quantity
cannot account for the shear stress and its evolution with strain. 
Indeed, the expression of the stress tensor suggests that   the useful 
information for the analysis of shear stress is the density and average force 
as a function of contact orientation. Such functions can be expanded in spherical 
harmonics in 3D (\cite{Ouadfel2001}).    

\begin{figure}
\centering
\includegraphics[width=8cm]{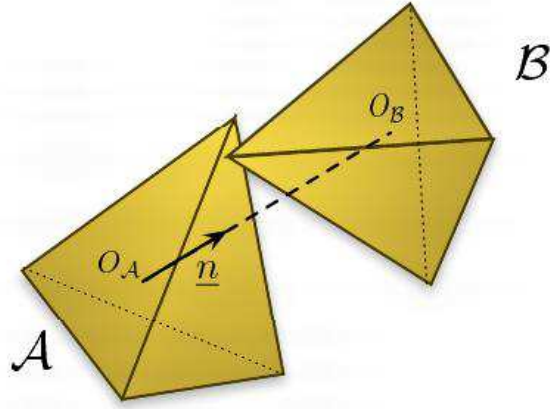}
\caption{Geometry of a contact between two polyhedra.}
\label{image_theta_lambda}    
\end{figure}

Let $\bm n$ be the unit vector along the branch vector   $\bm \ell$ ; Fig. \ref{image_theta_lambda}. We set 
\begin{equation}
\centering
\bm \ell =  \ell \bm n,
\end{equation}
where $\ell$ is the length of the branch vector. 
We remark that the unit vector $\bm n$ does not coincide with the contact 
normal except in the case of spheres.  
We consider the components  
of the contact force in a local frame defined by $\bm n$ and an orthoradial 
unit vector  $\bm t$:
\begin{equation}
\centering
\bm f = f_n \bm n + f_t \bm t,
\label{f=fn_3D}
\end{equation}
where $f_n$ and $f_t$ are the radial and orthoradial components of the 
contact force, respectively. The writing of Eq. (\ref{f=fn_3D}) assumes that 
$\bm t$ is oriented along the orthoradial force.    

\begin{figure}
\centering
\includegraphics[width=8cm]{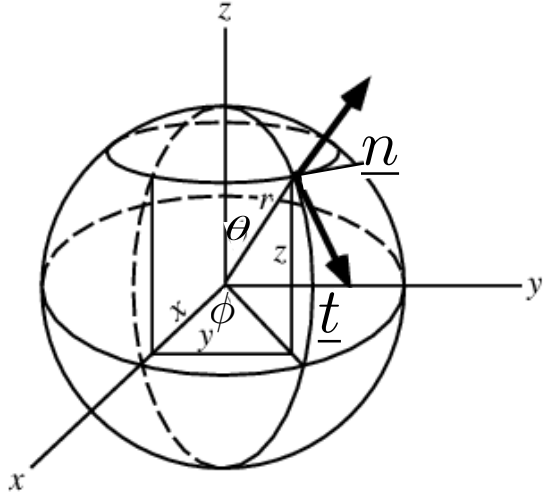}
\caption{Spherical coordinates.}
\label{harmonique_spherique}    
\end{figure}

We now define the angular averages associated with the branch vectors $\bm \ell$ and 
contact force vectors $\bm f$. Let ${\cal A}(\Omega)$ be the set of  
branch vectors pointing in the direction $\Omega \equiv (\theta,\phi)$ up to a 
solid angle   $d\Omega$ and $N_c(\Omega)$ its cardinal. 
The angles $\theta$ and $\phi$ are shown in Fig. \ref{harmonique_spherique}. 
The angular averages are defined as follows:
\begin{eqnarray}
P_\Omega (\Omega) &=& \frac{N_c(\Omega)}{N_c}, \\
\langle \ell \rangle (\Omega) &=& \frac{1}{N_c(\Omega)} \sum_{c\in {\cal A}(\Omega)} \ell^c, \\
\langle f_n \rangle (\Omega) &=& \frac{1}{N_c(\Omega)} \sum_{c\in {\cal A}(\Omega)} f_n^c, \\
\langle f_t \rangle (\Omega) &=& \frac{1}{N_c(\Omega)} \sum_{c\in {\cal A}(\Omega)} f_t^c, 
\label{eq:pom}
\end{eqnarray}
where $N_c = \int N_c(\Omega) d\Omega$ is the total number of contacts, and 
$ \ell^c$, $ f_n^c$, and $ f_t^c$ are the actual values of branch vector length, radial force 
and orthoradial force for contact $c$, respectively.  

\begin{figure}[htbp] 
  \centering
  \includegraphics[width=12cm]{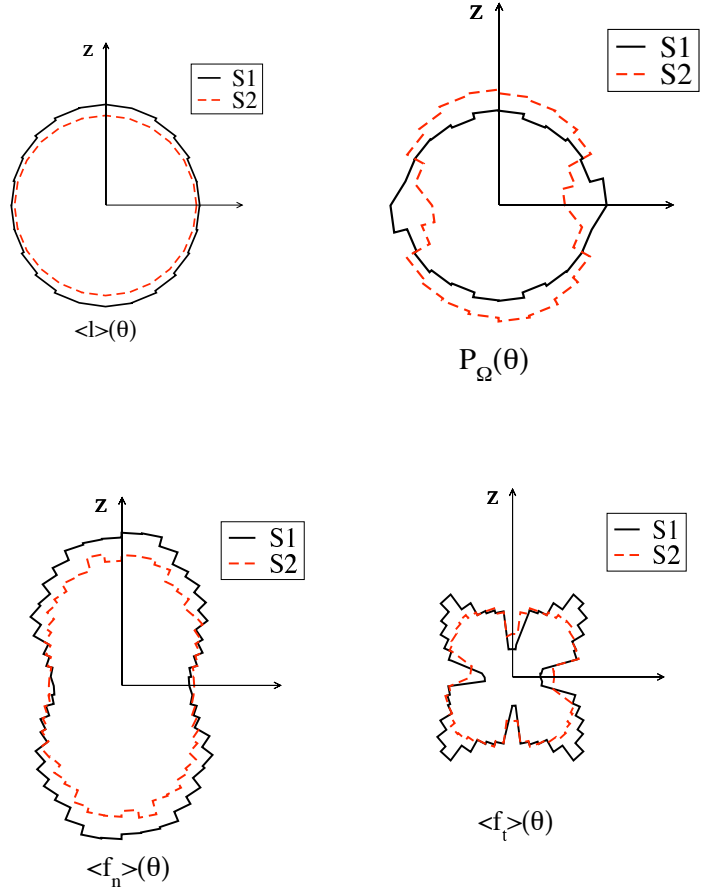}
\caption{Polar representation of density probability function $P_\Omega (\theta)$, $\langle f_n\rangle (\theta)$,
    $\langle f_t\rangle (\theta)$ and $\langle \ell \rangle (\theta)$  for S1 et S2 in residual state.}
\label{polar_branche_3D}    
\end{figure}

Under the axisymmetric conditions of our simulations, the four functions defined in 
Eq. (\ref{eq:pom}) are independent of $\phi$. Fig. \ref{polar_branche_3D} displays 
a polar representation of these functions in the $\theta$-plane for polyhedra (S1) and 
spheres (S2) at $\varepsilon_q = 0.3$. We observe an anisotropic behavior in all cases 
except in $\langle \ell \rangle  (\theta)$ for S2. A weak anisotropy can be seen for S1 in the latter case. 
The peak values occur along the compression axis except for 
$\langle f_t \rangle (\theta)$ in which the peaks are inclined at $\pi/4$ with 
respect to the vertical. The magnitude of anisotropy is larger for polyhedra compared to spheres 
except for $P_\Omega (\theta)$ which is weakly anisotropic for polyhedra. 

The simple shapes of the above functions suggest that harmonic 
approximation based on spherical harmonics at leading 
terms captures their anisotropies. There are 9 second-order 
basis functions  $Y_m^l (\theta,\phi)$. But only the functions compatible with 
the symmetries of the problem, namely independent with respect to $\phi$ and 
$\pi$-periodic as a function of $\theta$, are admissible. For 
$P_\Omega (\theta)$ as a scalar, and $\langle \ell \rangle  (\theta)$ 
and $\langle f_n \rangle  (\theta)$ as radial components of the vectors, the 
only admissible functions  are $Y_0^0 = 1$ and $Y_2^0= 3\cos^2 \theta -1$. For 
$\langle f_t \rangle  (\theta)$ as orthoradial component of a vector, 
the only function independent of $\phi$ and perpendicular to $Y_0^0 = 1$ and $Y_2^0= 3\cos^2 \theta -1$ is $\sin 2\theta $. Hence, within the harmonic model of 
fabric and force, we have 
\begin{eqnarray}
\centering
P_ \Omega(\theta) &=& \frac{1}{4\pi} \{ \ 1 + a \ [3\cos^2\theta  -1] \ \}, \label{eq:a1} \\
\label{eq:a1}
\langle \ell \rangle (\theta) &=& \ell_0 \{ \ 1 + a_l \ [3\cos^2\theta  -1]  \ \} \label{eq:a2} \\
\label{eq:a2}
\langle f_n\rangle (\theta) &=& f_0 \{ \ 1 + a_n \ [3\cos^2\theta  -1]  \ \}, \label{eq:a3} \\
\label{eq:a3}
\langle f_t\rangle(\theta) &=& f_0 \  a_t \ \sin 2[\theta ] , 
\label{eq:a4}
\end{eqnarray}
where $a$, $a_l$, $a_n$ and $a_t$ are the anisotropy parameters, $\ell_0$ is the mean branch vector length, and $f_0$ the mean force. The probability density function 
$ P_ \Omega(\theta)$ is normalized to 1 ($\int_{\cal S} P_ \Omega(\Omega) d \Omega =1$, where $\cal S$ is a sphere of unit radius). 
The values of the anisotropies  $a$, $a_l$, $a_n$ and $a_t$  can be calculated 
from generalized {\em fabric tensors} introduced in Appendix \ref{Fab_tensors}.  

\begin{figure}
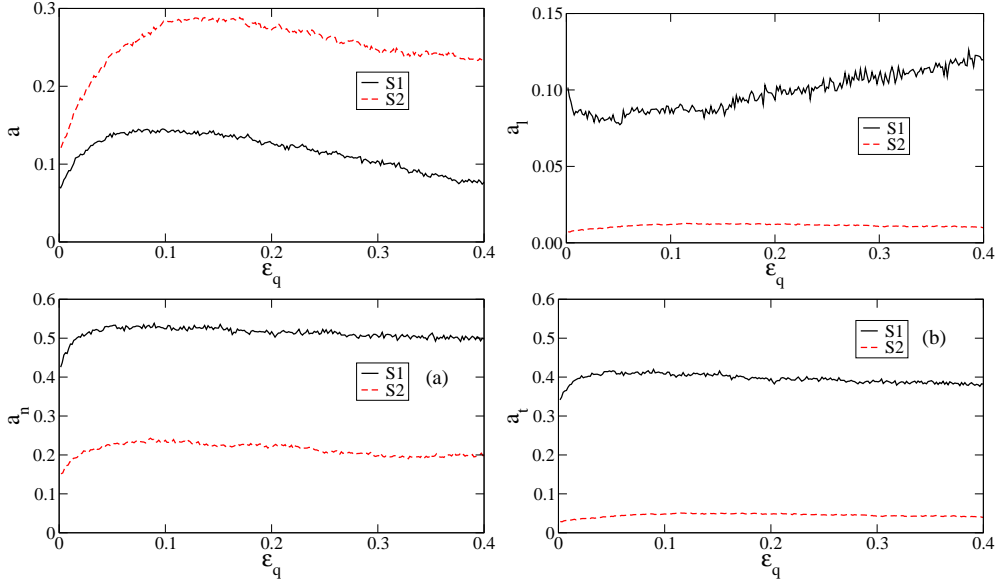

\centering
\includegraphics[width=6.5cm]{fig12a.eps}
\includegraphics[width= 6.5cm]{fig12b.eps}
\includegraphics[width= 6.5cm]{fig12c.eps}
\includegraphics[width= 6.5cm]{fig12d.eps}
\caption{Evolution of anisotropies  $a$, $a_l$, $a_n$ and $a_t$
    with $\varepsilon_q$ for packings S1 and S2.}
\label{anisol_epsilon_3D}    
\end{figure}

The evolution of the anisotropies with $\varepsilon_q$ are displayed in 
Fig.  \ref{anisol_epsilon_3D}  for our packings S1 and S2. 
The fabric orientation anisotropy $a$ increases with $\varepsilon_q$  
and relaxes to a plateau after passing by a pronounced peak. 
Its value is systematically larger for spheres than 
for polyhedra (by a factor 3 in the critical state). The branch vector anisotropy $a_l$ 
is quite low compared to other anisotropies and its value all along shearing is 
negligible for spheres. It is remarkable that $a_l$ for polyhedra 
declines (as $\varepsilon_p$, see Fig \ref{dilatance_all_3D}) 
at the beginning of shearing.  
The radial force anisotropy $a_n$ increases as the fabric anisotropy and tends 
to a plateau. But, in contrast to fabric anisotropy, its value is higher for polyhedra 
than for spheres. In other words, the aptitude of the polyhedra packing to 
develop large force anisotropy is correlated with particle shape rather than  
with fabric anisotropy (see section \ref{sec:net}). The orthoradial force anisotropy $a_t$ has a 
similar behavior except that it takes considerably higher 
values in the case of polyhedra compared to spheres.  
In the following section, we study the relationship between the 
fabric and force anisotropies.

%-----------------------------------------------------------------------------------------------------

\section{Origins of shear stress}
\label{sec:ori}

In this section, we analyze the stresses  in the framework of the  
harmonic approximation of granular microstructure introduced in the last section. 
Since this representation involves continuous functions of contact orientations, we 
need to express the stress tensor in integral form.  The stress tensor as defined 
in Eq. (\ref{eq:M}) is an average:
\begin{equation}  
\sigma_{\alpha \beta} = n_c \langle \ell^k_\alpha f^k_\beta \rangle_k,
\label{eq:sig}
\end{equation}
where $n_c = N_c / V$ is the number density of contacts, $\ell^k_\alpha$ is the 
$\alpha$ component of the branch vector at contact $k$ and  $\ell^k_\beta$ is the 
$\beta$ component of the force vector at contact $k$. The average is taken over all 
contacts $k$ in the control volume $V$. 
To express this mean as an integral,    
we introduce the joint probability density   
 $P_{\Omega f \ell}(\bm n, \bm f, \ell)$ of the force and branch 
 vectors (\cite{Bathurst1988,Rothenburg1989,Ouadfel2001}). 
Then, from Eq. (\ref{eq:sig}), we have          
\begin{equation}
\centering
\sigma_{\alpha\beta}= n_c   \int  P_{\Omega f \ell}(\bm n,\bm f, \ell) \ 
\ell(\bm n) \  f_\beta (\bm n, \ell) \  n_\alpha \   d\Omega \ d\bm f d\ell, 
\label{eq:sigint1}
\end{equation}
where $d\Omega = \sin \theta d\theta d\phi$. 

Equation (\ref{eq:sigint1}) can be simplified by integrating out the 
contribution of $\ell$. Assuming that $\bm f$ is independent of $\ell$ (an 
assumption which is verified with a good approximation), we get  
\begin{equation}
\centering
\sigma_{\alpha\beta}= n_c   \int  P_{\Omega f}(\bm n,\bm f) \ 
\langle \ell \rangle (\bm n) \  f_\beta (\bm n) \  n_\alpha \   d\Omega \ d\bm f, 
\label{eq:sigint2}
\end{equation}
where $\langle \ell \rangle \ P_{\Omega f}  = \int P_{\Omega f \ell} \ d\ell$. 

\begin{figure}
\centering
\includegraphics[width=8cm]{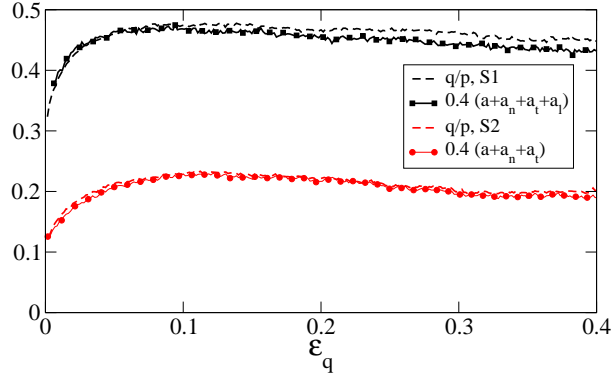}
\caption{The normalized
 shear stress $q/p$ as a function of shear strain $\varepsilon_q$ for the packings S1 and 
 S2 both from direct simulation data and theoretical prediction of  
 Eq. (\ref{eq_qp_3D_general}).}
\label{prediction_qp_3D}    
\end{figure}

Finally, integration of (\ref{eq:sigint2}) over force vector yields the 
following expression for the stress tensor:  
\begin{equation}
\centering
\sigma_{\alpha\beta}= n_c   \int  P_{\Omega}(\bm n) \ 
\langle \ell \rangle (\bm n) \ n_\alpha \  \langle f_\beta \rangle (\bm n) \    d\Omega, 
\label{eq:sigint3}
\end{equation}
where $\langle \bm f \rangle P_{\Omega}   = \int P_{\Omega f} \ d\bm f$. 
By introducing the average force components $\langle f_n \rangle $ and $\langle f_t \rangle $ 
in this equation, we get    
\begin{equation}
\centering
\sigma_{\alpha\beta} =  n_c  \int  P_ \Omega(\bm n) \ \langle \ell \rangle (\bm n) \ \{ \ \langle f_n \rangle (\bm n) \ n_\beta +  \langle f_t \rangle (\bm n) \ t_\beta \ \} d\Omega.
\label{eq:sigint4}
\end{equation}
This writing of the stress tensor involves the functions 
previously introduced with the 
harmonic representation of the fabric (Eqs. (\ref{eq:a1}), (\ref{eq:a2}), (\ref{eq:a3}) and  (\ref{eq:a4})). 
Inserting these functions in the integral expression (Eq. \ref{eq:sigint4}) and given the 
definitions of mean stress $p$ and stress deviator $q$ in Eq. (\ref{eq:q}), one 
gets  
\begin{eqnarray}
\centering
p & \simeq & n_c \ell_0 f_0, \\
\frac{q}{p} & \simeq & \frac{2}{5} \ (a + a_l+ a_n + a_t),
\label{eq_qp_3D_general}
\end{eqnarray}
where the cross products ($aa_l$, $aa_n$ and $aa_t$) among the anisotropies have been neglected. 
Our simulation data are in quantitative agreement with this 
``stress-force-fabric" relation (a term coined by Rothenburg and Bathurst in (\cite{Bathurst1988,Rothenburg1989}) )
both for spheres and polyhedra, as shown in Fig. \ref{prediction_qp_3D}, all 
along the shear.  
We note that the theoretical fit would have been less satisfactory for  polyhedra if the 
branch vector length anisotropy $a_l$  were omitted from the description.

Equation (\ref{eq_qp_3D_general}) is interesting as it exhibits the two origins of 
shear stress in a granular system: 1) the fabric anisotropies $a$ and $a_l$, related to 
the branch vector and 2) the force anisotropies $a_n$ and $a_t$, related to 
the contact force. Figure \ref{anisol_epsilon_3D} shows that the values of these  
anisotropy parameters underlying the shear stress depend on the particle shape. 
In particular, the total force anisotropy $a_n+ a_t$ compared to the total fabric anisotropy 
$a+a_l$ is much higher in the case of polyhedra. In the critical state, we have 
$a_n+ a_t \simeq 0.88$ and $a+a_l \simeq 0.2$ for 
polyhedra,  $a_n+ a_t \simeq 0.26$ and $a+a_l \simeq 0.24$ for spheres. The high value of 
the force anisotropy in the case of polyhedra comes from both radial and orthoradial 
components whereas in the sphere packing $a_t \simeq 0.05$ is much less important 
than $a_n \simeq 0.21$. This suggests that friction  is 
more directly involved in force transmission in the polyhedral packing than 
in the sphere packing. The strong contribution of force 
anisotropy to the polyhedra packing is a particle shape  effect related to the 
face-face contacts which carry most strong forces. This point will 
be analyzed in more detail below.            
 
%-----------------------------------------------------------------------------------------------------

\section{Force distributions}
\label{sec:for}

\begin{figure}
\centering
\includegraphics[width=6.5cm]{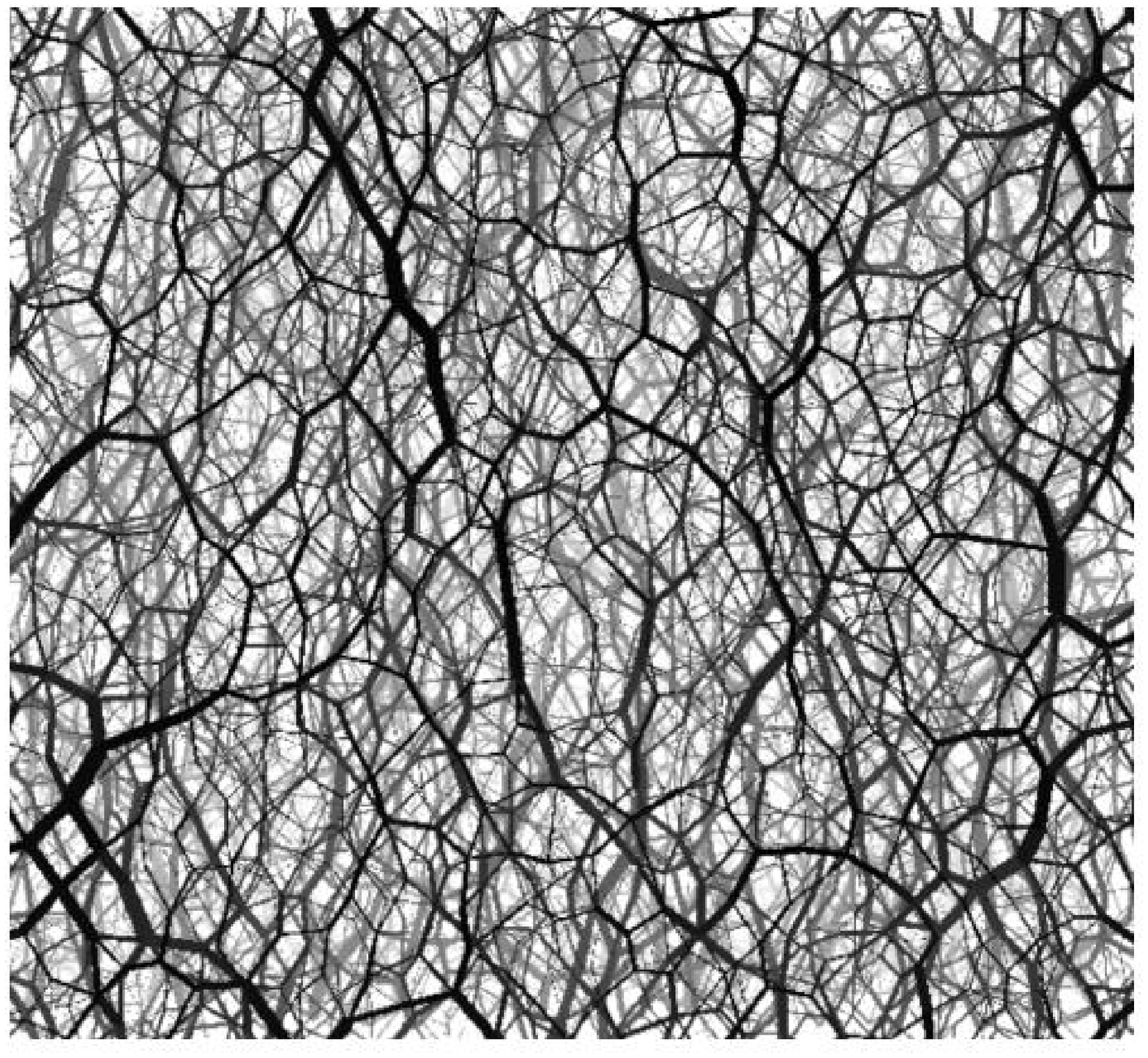}
\includegraphics[width=6.5cm]{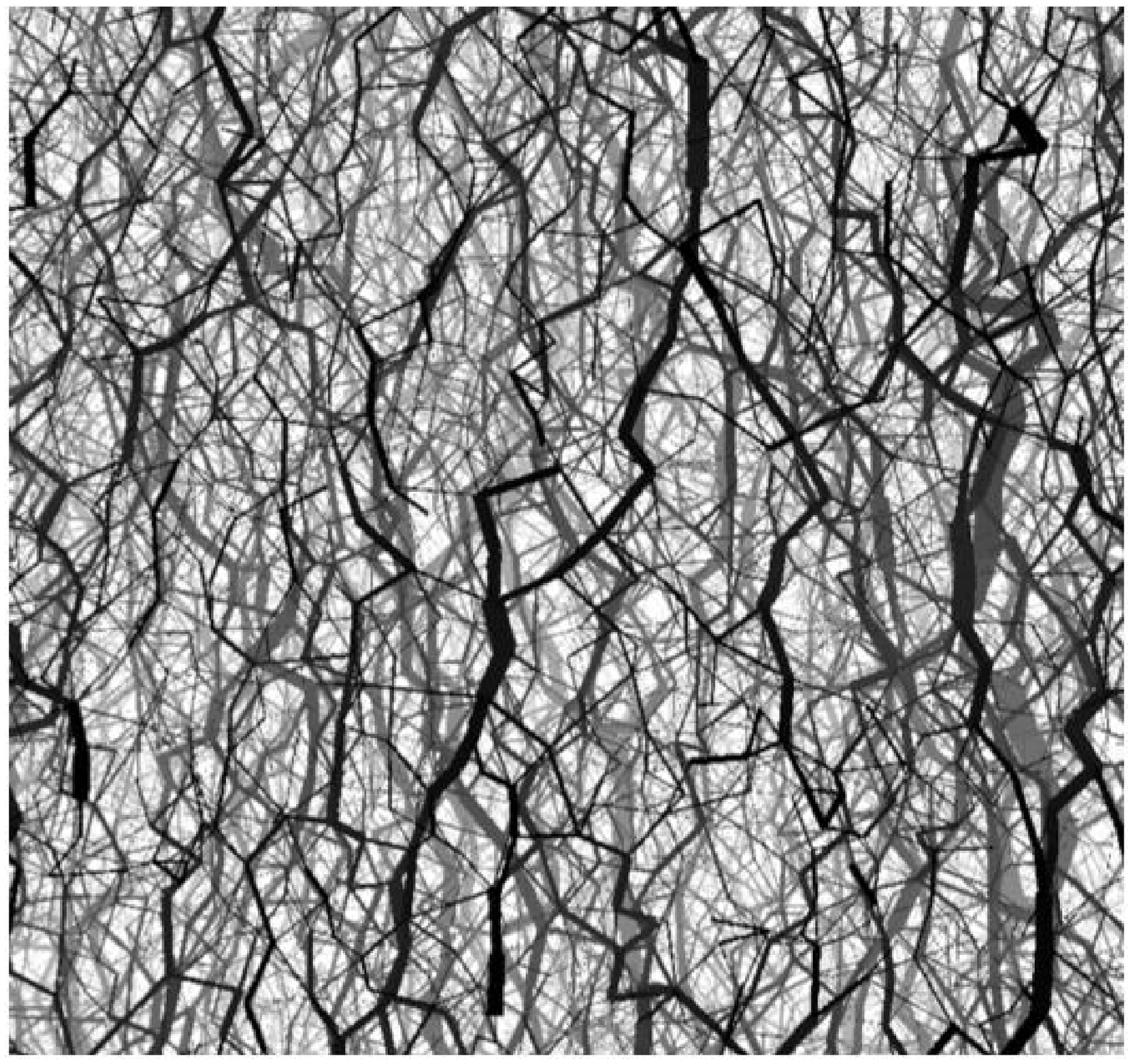}
\caption{Force maps in a portion of the packings S1 (right) and S2 (left). 
The segments are branch vectors with thickness proportional to the 
normal force,  and gray level 
proportional to the depth of field.}
\label{map_force_3D}    
\end{figure}

In this section, we study the probability density functions (pdf's)  $P(f_n)$ for sphere and 
polyhedra packings. Fig.  \ref{map_force_3D} shows typical maps of 
normal forces in a portion of both packings in the critical state.  
The 3D force chains can be observed in both packings, but they seem more 
tortuous in the case of polyhedra. 

\begin{figure}
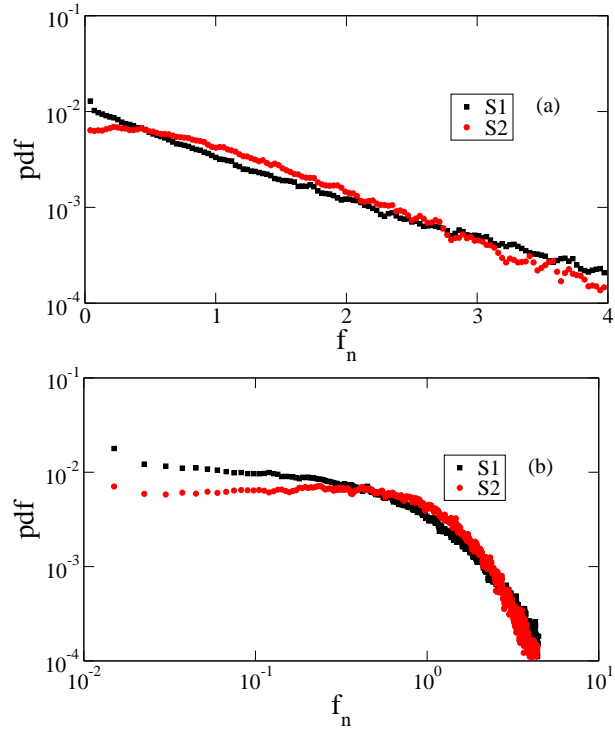

\centering
\includegraphics[width=8cm]{fig15a.eps}
\includegraphics[width=8cm]{fig15b.eps}
\caption{Probability density functions of normal forces in the 
packings of spheres and polyhedra.}
\label{pdf_3D}    
\end{figure}

The normal force pdf's are shown in Fig.  \ref{pdf_3D} on log-linear and log-log 
scales at $\varepsilon_q = 0,3$. In both pdf's, the strong forces, i.e. forces above the mean normal force $\langle f_n \rangle$, fall off exponentially: 
$P(f_n) \propto e^{- \beta  f_n / \langle f_n \rangle}$, with 
$\beta \simeq 0.9$ for S1 and $\beta \simeq 1.1$ for S2. In contrast, the shapes of the pdf's 
in the range of weak forces   
($f_n < \langle f_n\rangle$) are radically different. In the sphere packing, the pdf slightly 
bends down as $f_n \rightarrow 0$ but does not tend to zero. 
We observe also a small peak close to the mean force.  
This is consistent with several other numerical and experimental observations 
for isotropic packings (\cite{Lovol1999,Bardenhagen2000, Antony2001,Silbert2002,Metzger2004a,Majmudar2005}). 
In the case of polyhedra, the number of weak forces bends up as 
the force tends to zero. For both packings, the range of weak forces is well approximated by a  
power-law distribution :
\begin{equation}
\centering
\begin{array}{lcl}
P(f_n) \propto [\frac{f_n}{\langle f_n \rangle}]^{\alpha},  \\
\end{array}
\label{eqn:faible_3D}
\end{equation}      
with $\alpha=-0.24$ for S1 and $\alpha=0.05$ for S2. The divergence 
of the number of weak forces 
in S1 should be attributed to the polyhedral shape of the particles favoring the arching effect an hence 
a higher fraction of weak forces. The coefficient of friction has a similar effect though to a lesser extent. 
We find, however, that in both systems the fraction 
of weak forces   ($f_n < \langle f_n\rangle$) is 
about 60\%.

%-----------------------------------------------------------------------------------------------------

\section{Contact networks of polyhedral particles}
\label{sec:net}

In the case of the polyhedra packing, it is interesting to investigate the organization of the 
contact network in terms of simple, double and triple contacts. 
The respective fractions of these contact types and their contributions to 
the structural anisotropy and force transmission are the key quantities for understanding the 
effect of particle shape on the shear strength properties of granular media.  
In fact, one expects that the triple (face-to-face) contacts play an essential role in force 
transmission. This feature was observed in the case of polygon 
packings for side-to-side contacts (\cite{Azema2007}).  

\begin{figure}
\centering
\includegraphics[width=8cm]{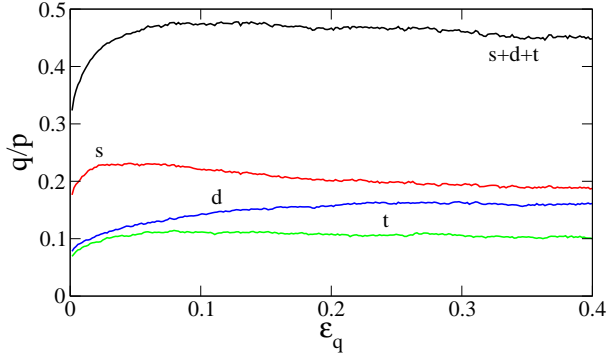}
\caption{Evolution of partial shear stresses as a function of shear strain 
for simple (s), double (d) and triple (t) contacts, as well as the total shear stress (s+d+t).}
\label{qp_sdt}    
\end{figure}

Considering the discrete expression of the stress tensor in Eq. (\ref{eq:M}) 
and restricting the summation to each contact type allows us to perform 
an additive decomposition: 
\begin{equation}
\centering
\bm \sigma = \bm \sigma_s + \bm \sigma_d + \bm \sigma_t,
\label{eq_sigma_sdt}
\end{equation}
where the subscripts $s$, $d$ and $t$ design the respective contributions of simple, double and triple contacts. The corresponding stress deviators $q_s$, $q_d$ and $q_t$ are then 
calculated and normalized by the mean stress $p$. Fig. \ref{qp_sdt} 
shows the evolution of partial shear stresses $q_s/p$, $q_d/p$ and $q_t/p$ 
as a function of shear strain $\varepsilon_q$. 
The contribution of simple contacts is larger for double and triple contacts. 
However, the double and triple contacts support together the largest 
portion of the overall shear stress, i.e. $q_d + q_t > q_s$.

\begin{figure}
\centering
\includegraphics[width=8cm]{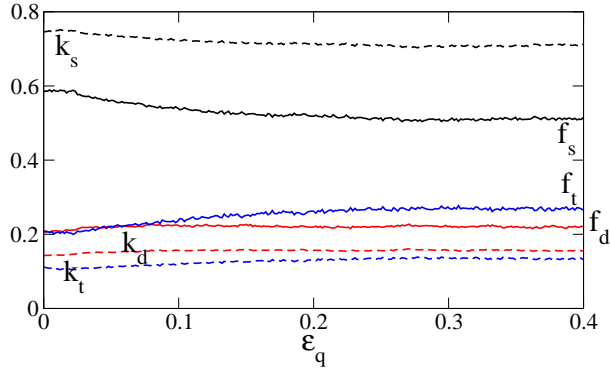}
\caption{Proportions $k_s$, $k_d$ and $k_t$ of simple, double and triple contacts 
(dashed lines), and the relative average forces $f_s$, $f_d$ and 
    $f_t$ (full lines) supported by each contact type as a function of shear strain 
     $\varepsilon_q$.}
\label{kskdkt}    
\end{figure}

The partial shear stress supported by each contact type  depends  
on both the number of its contacts and their   
mean force.  Fig. \ref{kskdkt} shows the proportions $k_s$, $k_d$ and $k_t$ 
of simple, double and triple contacts as a function of shear strain. 
$k_s$ declines during shear from 0.75 to 0.71 whereas  $k_d$ and $k_t$ increase 
from 0.14 to 0.15 and from 0.11 to 0.14, respectively.  
Hence, the critical state is characterized by  
$k_s \simeq 0.7$ et $k_t\simeq k_d\simeq0.15$. 
Fig.  \ref{kskdkt} also shows the {\em relative} mean forces $f_s$, $f_d$ and $f_t$ 
defined by 
\begin{eqnarray}
\centering
f_s &=& k_s \langle f_n \rangle_s / \langle f_n \rangle, \\
f_d &=& k_d \langle f_n \rangle_d / \langle f_n \rangle,\\
f_t &=& k_t \langle f_n \rangle_t / \langle f_n \rangle,
\end{eqnarray}
where $\langle f_n \rangle_s$, $\langle f_n \rangle_d$ and $\langle f_n \rangle_t$ 
correspond to the mean normal forces of simple, double and triple contacts. 
We see that $f_s$ declines slightly with strain but is nearly two times 
larger  than $f_t$ and 2.3 times larger than $f_d$ in the critical state. We have 
$f_s\simeq  f_d + f_t$. Hence, the lower contribution of 
triple contacts with respect to shear stress can be attributed to both 
the low level of the mean force ($f_t < 0.3$) sustained by this class and to their 
weak number ($< 15 \%$).   

Following the same procedure as for the stress tensor, we now 
perform a similar  decomposition of the fabric tensor  
$\bm F$, defined by Eq. (\ref{eq_all_tenseur_H}), into three terms: 
\begin{eqnarray}
\centering
\bm F &=& \bm F_s + \bm F_d + \bm F_t, \\
\label{eq_fabrique_sdt}
\end{eqnarray}
where $\bm F_s$, $\bm F_d$ and $\bm F_t$ are the contributions of simple, double and triple contacts. The corresponding anisotropies $a_s$, $a_d$ and $a_t$ can be extracted, but since the principal directions of these partial fabric tensors 
are not necessarily identical to those of the overall fabric tensor, we define the ``signed" anisotropies by multiplying each partial anisotropy $a_i$ by a phase factor 
$\cos 2 (\theta_F  - \theta_{F_i})$:   
\begin{eqnarray}
\centering
a'_i &=& a_i \cos 2(\theta_F-\theta_{F_i}).
\label{eq_aa'_sdt}
\end{eqnarray}

\begin{figure}
\centering
\includegraphics[width=8cm]{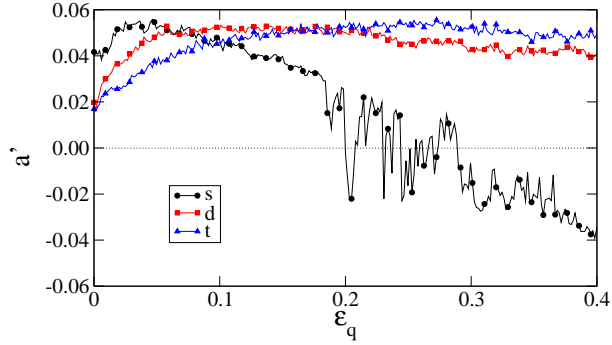}
\caption{Evolution of the signed anisotropies $a'$ of simple (s), double (d) and 
triple (t) contacts as a function of shear strain  $\varepsilon_q$.}
\label{b'sdt}    
\end{figure}

\begin{figure}
\centering
\includegraphics[width=8cm]{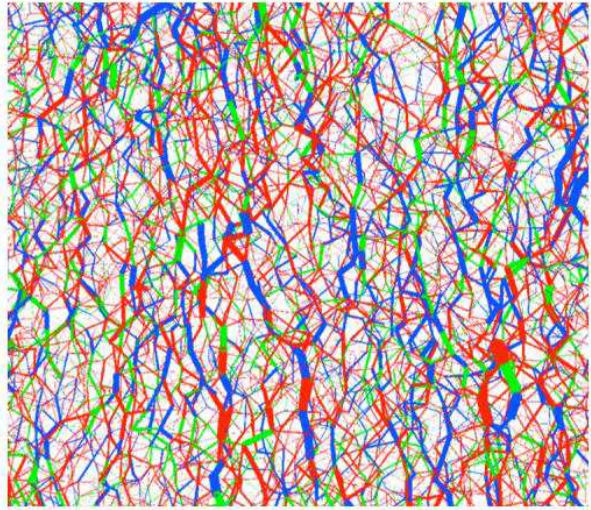}
\caption{Map of contact forces projected along branch vectors 
at  $\varepsilon_q=0.4$. Line thickness is proportional to the force.    
The simple, double and triple contacts are in red (dark gray), in 
green (light gray) and in blue (black). }
\label{map_sdt}    
\end{figure}

Figure  \ref{b'sdt} shows the evolution of signed anisotropies of the three contact classes. 
We see that $a'_d$ and $a'_t$ increase with shear strain and tend to the 
limit value $\simeq 0.04$. As to $a'_s$, we observe an initial increase followed by 
rapid decrease and change of sign  at $\varepsilon_q\simeq 0.2$. 
This evolution means that during shear the branch vectors of simple contacts tend  
to become increasingly perpendicular to the major principal 
direction (the direction of compression). A map of contact forces projected 
along the branch vectors is displayed in Fig. \ref{map_sdt} in different colors 
according to the type of contact. The triple contacts, despite their lower proportion, appear 
clearly to be correlated in the form of long  chains across the packing. These are mostly 
parallel to the direction of compression.  We also observe a large number of 
weak forces mainly at simple contacts.

\begin{figure}
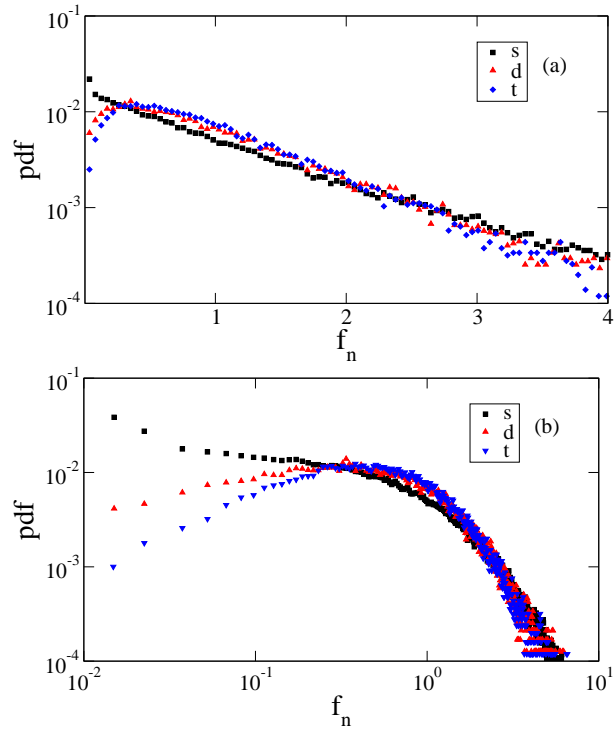

\centering
\includegraphics[width=8cm]{fig20a.eps}
\includegraphics[width=8cm]{fig20b.eps}
\caption{Probability distribution functions of radial forces at simple (s), double (d) and 
triple (t) contacts on log-linear (a) and log-log (b) scales.}
\label{pdf_sdt_3D}    
\end{figure}

The pdf's of normal forces are shown in Fig. \ref{pdf_sdt_3D} separately 
for simple, double and triple contacts. The three contact types are involved in 
strong and weak networks. The strong forces have in all cases an exponential 
behavior but a major difference is observed in the range of weak forces where the proportion 
of simple contacts prevails. This correlation between simple and weak contacts is 
interesting as it clearly reveals the contrast between simple 
contacts, on one hand, and double and triple contacts, on the other hand, in 
the organization of the force network.       

\begin{figure}
\centering
\includegraphics[width=8cm]{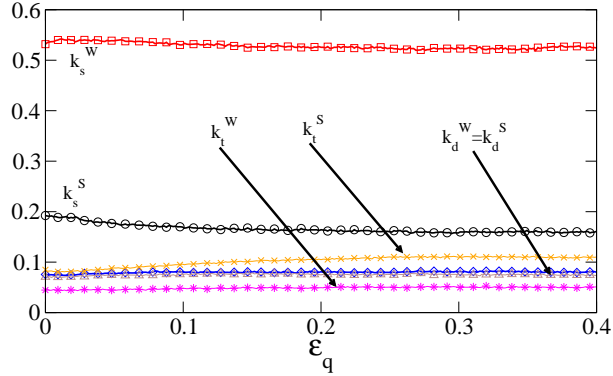}
\caption{Proportions $k_s^S$, $k_d^S$ and $k_t^S$  of simple (s), double (d) and 
triple (t) contacts in the strong network (S) and the corresponding proportions  
$k_s^W$, $k_d^W$ and $k_t^W$ in the weak network (W) as a function of shear strain.}
\label{kswksfkdwkds3D}    
\end{figure}

In order to situate the simple, double and triple contacts with respect to the force network, 
we have plotted in Fig. \ref{kswksfkdwkds3D}  the proportions 
$k_s^S$, $k_d^S$ et $k_t^S$ of the three contact sets in the strong network and 
the corresponding proportions  $k_s^W$, $k_d^W$ et $k_t^W$ in the weak network 
as a function of shear strain $\varepsilon_q$. It is interesting to note that the proportion 
of weak simple contacts is quite high ($\simeq 0.55$). The proportions  
 $k_d^W$ et $k_d^S$ of weak and strong double contacts 
 are identical  ($\simeq 0.07$). Finally, we see that  most 
 double contacts   belong to the strong network ($k_t^S \simeq 2 k_t^W$).

%-----------------------------------------------------------------------------------------------------

\section{Conclusion}
\label{sec:con}

In this paper, granular materials composed of irregular polyhedral 
particles were numerically investigated. Macroscopic and microstructural 
properties were analyzed by (1) direct comparison with a similar 
packing composed of spherical particles and (2) characterization of 
contact networks and force transmission. A novel finding of this work is 
that the origin of enhanced shear strength in a polyhedra packing compared to a sphere packing 
lies in force anisotropy induced by particle shape. The fabric anisotropy associated with  
the network of branch vectors is lower in the polyhedra packing. 
This finding extends the results of a previous study of pentagonal 
particles in two dimensions to three dimensions \cite{Azema2007}. In other words, 
the force anisotropy, partially underlying  shear strength,  
 is mainly controlled by the fabric anisotropy in a sphere packing. This mechanism 
breaks down to some extent in a packing of polyhedra where 
force anisotropy results mainly from the ``facetted" particle shape.   

The face-face contacts were shown to belong mostly 
to the strong force network. The local equilibrium structures involving 
face-face and edge-face contacts accommodate force lines that are 
basically unstable with spherical particles. Hence, the term 
``arching'' seems to be more adapted 
to the description of force patterns in an assembly of polyhedra  
than in an assembly of spheres. This effect is crucial for the 
probability density function of normal forces   
in the range of weak forces that is well approximated by 
a decreasing power-law in the case of polyhedra. 

In this investigation the polyhedra were irregular with a 
given number of faces, edges and vertices. These shape parameters 
can now be changed and the resulting packings can be analyzed 
along the same lines as  in the present investigation. 
Since  the face-face contacts seem to play a key role, 
it would be interesting to consider irregular polyhedra with 
less faces in number but with larger areas. 
From a mechanical point of view, there should be 
little difference between a small face and a vertex. The best 
shape from the shear strength viewpoint can be obtained with 
a large number of faces of large area, but these two conditions 
can not be realized at the same time. It seems thus 
that an optimal polyhedral shape should exist with a number of faces 
of not two low areas. The work is under way to elucidate 
this point and other aspects of the problem concerned 
with packing structure by systematically changing the particle shape parameters. 
    
We acknowledge assistance by F. Dubois with the LMGC90 platform 
employed for the simulations, as well as the precious help of V. Richefeu 
with 3D visualization of forces. This work was funded 
by the French Railway Society, the SNCF, and the R\'egion Languedoc-Roussillon 
of France.

%---------------------------------------------------------------------------
\appendix

\section{Fabric tensors}
\label{Fab_tensors}

The anisotropies $a$, $a_n$, $a_t$ and $a_l$ can be calculated from the 
tensors $\bm F$, $\bm H^{(n)}$,  $\bm H^{(t)}$ and 
 $\bm H^{(l)}$ defined by (\cite{Bathurst1988,Rothenburg1989,Ouadfel2001}) :  
\begin{eqnarray}
\centering
\bm F_{\alpha \beta}          &=&   \int_\mathcal{S} P_\Omega (\theta) \ n_\alpha n_\beta \  d\Omega,\\
\bm H^{(n)}_{\alpha \beta} &=&  \int_\mathcal{S} \langle f_n \rangle (\theta) \ n_\alpha n_\beta \  d\Omega,\\
\bm H^{(t)}_{\alpha \beta}  &=&  \int_\mathcal{S} \langle f_t \rangle (\theta) \ n_\alpha t_\beta \  d\Omega,\\
\bm H^{(l)}_{\alpha \beta}  &=&  \int_\mathcal{S} \langle \ell \rangle (\theta) \ n_\alpha n_\beta \  d\Omega.
\label{eq_all_tenseur_H}
\end{eqnarray}
Using the equations (\ref{eq:a1}), (\ref{eq:a2}), (\ref{eq:a3}) and  (\ref{eq:a4}), it is 
then easy to show that the corresponding anisotropies are :
\begin{eqnarray}
\centering
a &=& \frac{5}{2 } \ \frac{F_3 - F_1}{ tr \bm F}, \\
a_n &=& \frac{5}{2} \  \frac{H^{(n)}_3 - H^{(n)}_1}{ tr \bm H^{(n)}}, \\
a_t &=& \frac{5}{2} \    \frac{H^{(t)}_3 - H^{(t)}_1}{tr \bm H^{(n)}}, \\
a_l &=& \frac{5}{2} \ \frac{H^{(l)}_3 - H^{(l)}_1}{tr \bm H^{(l)}}, 
\end{eqnarray}
where $tr \bm H^{(n)} = \langle f \rangle $,  $tr \bm F =1$ et $tr \bm H^{(l)} = \ell_0 $ .

\bibliographystyle{plainnat}
\bibliography{emilien}

\begin{thebibliography}{49}
\providecommand{\natexlab}[1]{#1}
\providecommand{\url}[1]{\texttt{#1}}
\expandafter\ifx\csname urlstyle\endcsname\relax
  \providecommand{\doi}[1]{doi: #1}\else
  \providecommand{\doi}{doi: \begingroup \urlstyle{rm}\Url}\fi

\bibitem[Airey and Wood(1988)]{Airey1988}
D.W. Airey and D.M. Wood.
\newblock \emph{"The Cambridge true triaxial apparatus", Advanced Triaxial
  Testing of Soil and Rock}.
\newblock Rebert T. Donaghe, Ronald C. Chaney \& Marshall L. Silver, 1988.

\bibitem[Alonso-Marroquin and Herrmann(2002)]{Alonso-Marroquin2002}
F.~Alonso-Marroquin and H.~J. Herrmann.
\newblock Calculation of the incremental stress-strain relation of a polygonal
  packing.
\newblock \emph{Phys. Rev. E}, 66\penalty0 (2):\penalty0 021301--, August 2002.

\bibitem[Antony(2001)]{Antony2001}
S.~J. Antony.
\newblock Evolution of force distribution in three-dimensional granular media.
\newblock \emph{Phys Rev E}, 63:\penalty0 011302, 2001.

\bibitem[Antony and Kuhn(2004)]{Antony2004}
S.J Antony and M.R. Kuhn.
\newblock Influence of particle shape on granular contact signatures and shear
  strength: new insights from simulations.
\newblock \emph{International Journal of Solids and Structures}, 41\penalty0
  (21):\penalty0 5863--5870, October 2004.

\bibitem[Az\'ema et~al.(2007)Az\'ema, Radjai, Peyroux, and Saussine]{Azema2007}
E.~Az\'ema, F.~Radjai, R.~Peyroux, and G.~Saussine.
\newblock Force transmission in a packing of pentagonal particles.
\newblock \emph{Phys. Rev. E}, 76:\penalty0 011301, 2007.

\bibitem[Bardenhagen et~al.(2000)Bardenhagen, Brackbill, and
  Sulsky]{Bardenhagen2000}
S.~G. Bardenhagen, J.~U. Brackbill, and D.~Sulsky.
\newblock Numerical study of stress distribution in sheared granular material
  in two dimensions.
\newblock \emph{Phys. Rev. E}, 62:\penalty0 3882--3890, 2000.

\bibitem[Bathurst and Rothenburg(1988)]{Bathurst1988}
R.~J. Bathurst and L.~Rothenburg.
\newblock Micromechanical aspects of isotropic granular assemblies with linear
  contact interactions.
\newblock \emph{J. Appl. Mech.}, 55:\penalty0 17, 1988.

\bibitem[Cambou et~al.(2004)Cambou, Dubujet, and Nouguier-Lehon]{Cambou2004}
B.~Cambou, Ph. Dubujet, and C.~Nouguier-Lehon.
\newblock Anisotropy in granular materials at different scales.
\newblock \emph{Mechanics of Materials}, 36\penalty0 (12):\penalty0 1185--1194,
  December 2004.

\bibitem[Coppersmith et~al.(1996)Coppersmith, Liu, Majumdar, Narayan, and
  Witten]{Coppersmith1996}
S.~N. Coppersmith, {C.-h.} Liu, S.~Majumdar, O.~Narayan, and T.~A. Witten.
\newblock Model for force fluctuations in bead packs.
\newblock \emph{Phys. Rev. E}, 53\penalty0 (5):\penalty0 4673--4685, 1996.

\bibitem[Cundall and Strack(1979)]{Cundall1979}
P.~A. Cundall and O.D.L. Strack.
\newblock Discrete numerical model for granular assemblies.
\newblock \emph{geotechnique}, 29\penalty0 (1):\penalty0 47--65, 1979.

\bibitem[Cundall(1988)]{Cundall1988}
P.A. Cundall.
\newblock Formulation of a three-dimensionnal distinct element model-part i: a
  scheme to detect and represent contacts in a system composed of many
  polyhedral blocks.
\newblock \emph{Int. J. Rock Mech. Min Sci \& Geomech. Abstr.}, 1988.

\bibitem[Dubois and Jean(2003)]{DUBOIS2003}
F.~Dubois and M.~Jean.
\newblock Lmgc90 une plateforme de d\'eveloppement d\'edi\'ee \`a la
  mod\'elisation des probl\`emes d'int\'eraction.
\newblock In \emph{Actes du sixi\`eme colloque national en calcul des
  structures - CSMA-AFM-LMS -}, volume~1, pages 111--118, 2003.

\bibitem[GDR-MiDi(2004)]{GDR-MiDi2004}
GDR-MiDi.
\newblock On dense granular flows.
\newblock \emph{Eur. Phys. J. E}, 14:\penalty0 341--365, 2004.

\bibitem[Jean and Moreau(1992)]{jean1992}
M.~Jean and J.~J. Moreau.
\newblock Unilaterality and dry friction in the dynamics of rigid body
  collections.
\newblock In \emph{Proceedings of Contact Mechanics International Symposium},
  pages 31--48, Lausanne, Switzerland, 1992. Presses Polytechniques et
  Universitaires Romandes.

\bibitem[Kruyt and Rothenburg(1996)]{Kruyt1996}
N.~P. Kruyt and L.~Rothenburg.
\newblock Micromechanical definition of strain tensor for granular materials.
\newblock \emph{ASME Journal of Applied Mechanics}, 118:\penalty0 706--711,
  1996.

\bibitem[Kruyt and Rothenburg(2004)]{Kruyt2004}
N.~P. Kruyt and L.~Rothenburg.
\newblock Kinematic and static assumptions for homogenization in micromechanics
  of granular materials.
\newblock \emph{Mechanics of Materials}, 36\penalty0 (12):\penalty0 1157--1173,
  December 2004.

\bibitem[Lim and MacDowel(2005)]{Lim2005}
W.L. Lim and G.R. MacDowel.
\newblock Discrete element modelling of railway ballast discrete element
  modelling of railway ballast.
\newblock \emph{Granular Matter}, 7:\penalty0 19--29, 2005.

\bibitem[Liu et~al.(1995)Liu, Nagel, Schecter, Coppersmith, Majumdar, Narayan,
  and Witten]{Liu1995a}
{C.-h.} Liu, S.~R. Nagel, D.~A. Schecter, S.~N. Coppersmith, S.~Majumdar,
  O.~Narayan, and T.~A. Witten.
\newblock Force fluctuations in bead packs.
\newblock \emph{Science}, 269:\penalty0 513, 1995.

\bibitem[Lobo-Guerrero and Vallejo(2006)]{Lobo-Guerrero2006}
S.~Lobo-Guerrero and L.~E. Vallejo.
\newblock Discrete element method analysis of railtrack ballast degradation
  during cyclic loading.
\newblock \emph{Granular Matter}, 8:\penalty0 195--2004, 2006.

\bibitem[Lovol et~al.(1999)Lovol, Maloy, and Flekkoy]{Lovol1999}
G.~Lovol, K.~Maloy, and E.~Flekkoy.
\newblock Force measurments on static granular materials.
\newblock \emph{Phys. Rev. E}, 60:\penalty0 5872--5878, 1999.

\bibitem[Lu and McDowel(2007)]{Lu2007}
M.~Lu and G.R. McDowel.
\newblock The importance of modelling ballast particle shape in the discrete
  element method.
\newblock \emph{Granular Matter}, 9:\penalty0 69--80, 2007.

\bibitem[Majmudar and Behringer(2005)]{Majmudar2005}
T.~S. Majmudar and R.~P. Behringer.
\newblock Contact force measurements and stresse-induced anisotropy in granular
  materials.
\newblock \emph{Nature}, 435:\penalty0 1079--1082, 2005.

\bibitem[Markland(1981)]{Markland1981}
J.G.D Morgan~E. Markland.
\newblock The effect of vibration on ballast beds.
\newblock \emph{Geotechnique}, 31\penalty0 (3):\penalty0 3,367--386, 1981.

\bibitem[Metzger(2004)]{Metzger2004a}
Philip~T. Metzger.
\newblock Granular contact force density of states and entropy in a modified
  edwards ensemble.
\newblock \emph{Phys. Rev. E}, 70\penalty0 (5 Pt 1):\penalty0 051303, Nov 2004.

\bibitem[Mitchell and Soga(2005)]{Mitchell2005}
J.K. Mitchell and K.~Soga.
\newblock \emph{Fundamentals of Soil Behavior}.
\newblock Wiley, NY, 2005.

\bibitem[Moreau(1997)]{Moreau1997}
J.~J. Moreau.
\newblock Numerical investigation of shear zones in granular materials.
\newblock In D.~E. Wolf and P.~Grassberger, editors, \emph{Friction, Arching,
  Contact Dynamics}, pages 233--247, Singapore, 1997. World Scientific.

\bibitem[Moreau(1994)]{Moreau1994}
J.J. Moreau.
\newblock Some numerical methods in multibody dynamics : application to
  granular.
\newblock \emph{Eur. J. Mech. A/Solids}, 13:\penalty0 93--114, 1994.

\bibitem[Mueth et~al.(1998)Mueth, Jaeger, and Nagel]{Mueth1998a}
D.~M. Mueth, H.~M. Jaeger, and S.~R. Nagel.
\newblock Force distribution in a granular medium.
\newblock \emph{Phys. Rev. E.}, 57\penalty0 (3):\penalty0 3164--3169, 1998.

\bibitem[Nezami et~al.(2004)Nezami, Hashash, Zaho, and Ghaboussi]{nezami2004}
E.G. Nezami, Y.M.A Hashash, D.~Zaho, and J.~Ghaboussi.
\newblock A fast contact detection for 3-d discrete element method.
\newblock \emph{Computers and Geotechnics}, 31:\penalty0 575--587, 2004.

\bibitem[Nezami et~al.(2006)Nezami, Hashash, Zaho, and Ghaboussi]{Nezami2006}
E.G. Nezami, Y.M.A Hashash, D.~Zaho, and J.~Ghaboussi.
\newblock Shortest link method for contact detection in discrete element
  method.
\newblock \emph{Int. J. Numer. Anal. Meth. Geomech.}, 30:\penalty0 783--801,
  2006.

\bibitem[Nouguier-Lehon et~al.(2003)Nouguier-Lehon, Cambou, and
  Vincens]{Nouguier-Lehon2003}
C.~Nouguier-Lehon, B.~Cambou, and E.~Vincens.
\newblock Influence of particle shape and angularity on the behavior of
  granular materials: a numerical analysis.
\newblock \emph{Int. J. Numer. Anal. Meth. Geomech}, 27:\penalty0 1207--1226,
  2003.

\bibitem[Ouadfel and Rothenburg(2001)]{Ouadfel2001}
H.~Ouadfel and L.~Rothenburg.
\newblock `stress-force-fabric' relationship for assemblies of ellipsoids.
\newblock \emph{Mechanics of Materials}, 33\penalty0 (4):\penalty0 201--221,
  April 2001.

\bibitem[Pena et~al.(2005)Pena, Herrmann, Lizcano, and
  Alonso-Marroquin]{Pena2005}
A.A Pena, H.~J. Herrmann, A.~Lizcano, and F.~Alonso-Marroquin.
\newblock Investigation of the asymptotic states of granular materials using a
  discrete model of anisotropic particles.
\newblock In \emph{Powders and Grains 2005}, pages 697--700. A. A. Balkema,
  2005.

\bibitem[Pena et~al.(2006{\natexlab{a}})Pena, Garcia-Rojo, and
  Herrmann]{Pena2006}
A.A. Pena, R.~Garcia-Rojo, and H.J. Herrmann.
\newblock Influence of particle shape on sheared dense granular media.
\newblock \emph{Granular Matter}, In Press, 2006{\natexlab{a}}.

\bibitem[Pena et~al.(2006{\natexlab{b}})Pena, Lizcano, Alonso-Marroquin, and
  Herrman]{Pena2006a}
A.A. Pena, A.~Lizcano, F.~Alonso-Marroquin, and H.J. Herrman.
\newblock Fluctuations at the critical state of a polygonal packing.
\newblock \emph{Int. J. For Numer. Anal. Meth. Geomech.}, 00:\penalty0 1--12,
  2006{\natexlab{b}}.

\bibitem[P\'erales(2007)]{Perales2007}
R.~P\'erales.
\newblock \emph{Contribution \`a la mod\'elisation des structures maconn\'ees
  par approche discrete. Int\'egration vers une application industrielle}.
\newblock PhD thesis, Universit\'e Montpellier II (en cours), 2007.

\bibitem[Radjai and Roux(1999)]{Radjai1999}
F.~Radjai and S.~Roux.
\newblock Etats internes des milieux granulaires denses.
\newblock In \emph{14e Congres Francais de M\'ecanique. Toulouse}, 1999.

\bibitem[Radjai and Roux(2004)]{Radjai2004}
F.~Radjai and S.~Roux.
\newblock Contact dynamics study of 2d granular media : Critical states and
  relevant internal variables.
\newblock In H.~Hinrichsen and D.~E. Wolf, editors, \emph{The Physics of
  Granular Media}, pages 165--186, Weinheim, 2004. Wiley-VCH.

\bibitem[Radjai et~al.(1996)Radjai, Jean, Moreau, and Roux]{Radjai1996}
F.~Radjai, M.~Jean, J.J. Moreau, and S.~Roux.
\newblock Force distributions in dense two dimensional granular systems.
\newblock \emph{Phys. Rev. Letter}, 77:\penalty0 274--277, 1996.

\bibitem[Radjai et~al.(1998)Radjai, Wolf, Jean, and Moreau]{Radjai1998}
F.~Radjai, D.~E. Wolf, M.~Jean, and J.J. Moreau.
\newblock Bimodal character of stress transmission in granular packings.
\newblock \emph{Phys. Rev. Letter}, 80:\penalty0 61--64, 1998.

\bibitem[Rothenburg and Bathurst(1989)]{Rothenburg1989}
L.~Rothenburg and R.~J. Bathurst.
\newblock Analytical study of induced anisotropy in idealized granular
  materials.
\newblock \emph{Geotechnique}, 39:\penalty0 601--614, 1989.

\bibitem[Saussine(2004)]{Saussineoctober2004}
G.~Saussine.
\newblock \emph{Contribution \`a la mod\'elisation de granulats
  tridimensionnels : application au ballast}.
\newblock PhD thesis, Universit\'e Montpellier II, 2004.

\bibitem[Saussine et~al.(2006)Saussine, Cholet, Gautier, Dubois, Bohatier, and
  Moreau]{Saussine2006}
G.~Saussine, C.~Cholet, P.E. Gautier, F.~Dubois, C.~Bohatier, and J.J. Moreau.
\newblock Modelling ballast behaviour under dynamic loading. part1 : A 2d
  polygonal discrete element method approach.
\newblock \emph{Comput. Methods Appl. Mech. Eng.}, 195:\penalty0 2841 -- 2859,
  2006.

\bibitem[Silbert et~al.(2002)Silbert, Grest, and Landry]{Silbert2002}
L.~E. Silbert, G.~S. Grest, and J.~W. Landry.
\newblock Statistics of the contact network in frictional and frictionless
  granular packings.
\newblock \emph{Phys. Rev. E}, 66:\penalty0 1--9, 2002.

\bibitem[Staron and Radjai(2005)]{Staron2005}
L.~Staron and F.~Radjai.
\newblock Friction versus texture at the approach of a granular avalanche.
\newblock \emph{Phys. Rev. E}, 72:\penalty0 1--5, 2005.

\bibitem[Troadec(2002)]{Troadec2002a}
H.~Troadec.
\newblock \emph{Texture locale et plasticit\'e des mat\'eriaux granulaires}.
\newblock PhD thesis, Universit\'e Montpellier II, 2002.

\bibitem[Troadec et~al.(2002)Troadec, Radjai, Roux, and Charmet]{Troadec2002}
H.~Troadec, F.~Radjai, S.~Roux, and J.-C. Charmet.
\newblock Model for granular texture with steric exclusions.
\newblock \emph{Phys. Rev. E}, 66:\penalty0 041305, 2002.

\bibitem[Wood(1990)]{Wood1990}
D.M. Wood.
\newblock \emph{Soil behaviour and critical state soil mechanics}.
\newblock Cambridge University Press, Cambridge, England, 1990.

\bibitem[Wu and Thompson(2000)]{Wu2000}
Wu and Thompson.
\newblock The vibration behavior of railway track at high frequencies under
  multiple preloads and wheel interactions.
\newblock \emph{J Acoust Soc Am}, 108\penalty0 (3 Pt 1):\penalty0 1046--1053,
  Sep 2000.

\end{thebibliography}

\end{document}